\documentclass[twocolumn]{aastex62}
\usepackage{amsmath}
\usepackage{graphicx}
\usepackage{booktabs}

\newcommand{\zphot}{z_{\rm phot}}
\newcommand{\zspec}{z_{\rm spec}}
\newcommand{\alpc}{\alpha_{\lambda}}
\defcitealias{McGreer18}{M18}

\shorttitle{Discovery of $z\sim5$ Quasars with a Medium-band-based Approach}
\shortauthors{Kim et al.}

\begin{document}

\title{The Infrared Medium-deep Survey. VI. \\
Discovery of Faint Quasars at $z\sim5$ with a Medium-band-based Approach}

\correspondingauthor{Myungshin Im}
\email{yjkim@astro.snu.ac.kr, mim@astro.snu.ac.kr}

\author[0000-0003-1647-3286]{Yongjung Kim}
\affiliation{Center for the Exploration of the Origin of the Universe (CEOU), Building 45, Seoul National University, 1 Gwanak-ro, Gwanak-gu, Seoul 08826, Republic of Korea}
\affiliation{Astronomy Program, FPRD, Department of Physics \& Astronomy, Seoul National University, 1 Gwanak-ro, Gwanak-gu, Seoul 08826, Republic of Korea}

\author[0000-0002-8537-6714]{Myungshin Im}
\affiliation{Center for the Exploration of the Origin of the Universe (CEOU), Building 45, Seoul National University, 1 Gwanak-ro, Gwanak-gu, Seoul 08826, Republic of Korea}
\affiliation{Astronomy Program, FPRD, Department of Physics \& Astronomy, Seoul National University, 1 Gwanak-ro, Gwanak-gu, Seoul 08826, Republic of Korea}

\author[0000-0003-4847-7492]{Yiseul Jeon}
\affiliation{Center for the Exploration of the Origin of the Universe (CEOU), Building 45, Seoul National University, 1 Gwanak-ro, Gwanak-gu, Seoul 08826, Republic of Korea}
\affiliation{FEROKA Inc., 311-1, 108 Gasandigital2-ro, Geumcheon-gu, Seoul, Republic of Korea}

\author[0000-0002-3560-0781]{Minjin Kim}
\affiliation{Korea Astronomy and Space Science Institute, Daejeon 34055, Republic of Korea}
\affiliation{Department of Astronomy and Atmospheric Sciences, College of Natural Sciences, Kyungpook National University, Daegu 41566, Republic of Korea}

\author[0000-0002-2548-238X]{Soojong Pak}
\affiliation{Center for the Exploration of the Origin of the Universe (CEOU), Building 45, Seoul National University, 1 Gwanak-ro, Gwanak-gu, Seoul 08826, Republic of Korea}
\affiliation{School of Space Research, Kyung Hee University, 1732 Deogyeong-daero, Giheung-gu, Yongin-si, Gyeonggi-do 17104, Republic of Korea}

\author{Yoon Chan Taak}
\affiliation{Center for the Exploration of the Origin of the Universe (CEOU), Building 45, Seoul National University, 1 Gwanak-ro, Gwanak-gu, Seoul 08826, Republic of Korea}
\affiliation{Astronomy Program, FPRD, Department of Physics \& Astronomy, Seoul National University, 1 Gwanak-ro, Gwanak-gu, Seoul 08826, Republic of Korea}

\author{Changsu Choi}
\affiliation{Center for the Exploration of the Origin of the Universe (CEOU), Building 45, Seoul National University, 1 Gwanak-ro, Gwanak-gu, Seoul 08826, Republic of Korea}
\affiliation{Astronomy Program, FPRD, Department of Physics \& Astronomy, Seoul National University, 1 Gwanak-ro, Gwanak-gu, Seoul 08826, Republic of Korea}

\author{Jueun Hong}
\affiliation{Center for the Exploration of the Origin of the Universe (CEOU), Building 45, Seoul National University, 1 Gwanak-ro, Gwanak-gu, Seoul 08826, Republic of Korea}
\affiliation{Astronomy Program, FPRD, Department of Physics \& Astronomy, Seoul National University, 1 Gwanak-ro, Gwanak-gu, Seoul 08826, Republic of Korea}

\author{Minhee Hyun}
\affiliation{Center for the Exploration of the Origin of the Universe (CEOU), Building 45, Seoul National University, 1 Gwanak-ro, Gwanak-gu, Seoul 08826, Republic of Korea}
\affiliation{Astronomy Program, FPRD, Department of Physics \& Astronomy, Seoul National University, 1 Gwanak-ro, Gwanak-gu, Seoul 08826, Republic of Korea}

\author{Tae-Geun Ji}
\affiliation{Center for the Exploration of the Origin of the Universe (CEOU), Building 45, Seoul National University, 1 Gwanak-ro, Gwanak-gu, Seoul 08826, Republic of Korea}
\affiliation{School of Space Research, Kyung Hee University, 1732 Deogyeong-daero, Giheung-gu, Yongin-si, Gyeonggi-do 17104, Republic of Korea}

\author[0000-0003-1470-5901]{Hyunsung David Jun}
\affiliation{Korea Institute for Advanced Study, 85 Hoegi-ro, Dongdaemun-gu, Seoul 02455, Republic of Korea}

\author[0000-0002-8858-3188]{Marios Karouzos}
\affiliation{Nature Astronomy, Springer Nature, 4 Crinan street, London N1 9XW, UK}

\author[0000-0002-6925-4821]{Dohyeong Kim}
\affiliation{Kavli Institute for Astronomy and Astrophysics, Peking University, 5 Yi He Yuan Road, Haidian District, Beijing 100871, P. R. China}
\affiliation{Center for the Exploration of the Origin of the Universe (CEOU), Building 45, Seoul National University, 1 Gwanak-ro, Gwanak-gu, Seoul 08826, Republic of Korea}
\affiliation{Astronomy Program, FPRD, Department of Physics \& Astronomy, Seoul National University, 1 Gwanak-ro, Gwanak-gu, Seoul 08826, Republic of Korea}

\author[0000-0001-5120-0158]{Duho Kim}
\affiliation{Arizona State University, School of Earth and Space Exploration, P.O. Box 871404, Tempe, AZ 85287-1404, USA}

\author[0000-0002-1710-4442]{Jae-Woo Kim}
\affiliation{Korea Astronomy and Space Science Institute, Daejeon 34055, Republic of Korea}

\author[0000-0002-1418-3309]{Ji Hoon Kim}
\affiliation{Subaru Telescope, National Astronomical Observatory of Japan, 650 North A'ohoku Place, Hilo, HI 96720, USA}

\author{Hye-In Lee}
\affiliation{Center for the Exploration of the Origin of the Universe (CEOU), Building 45, Seoul National University, 1 Gwanak-ro, Gwanak-gu, Seoul 08826, Republic of Korea}
\affiliation{School of Space Research, Kyung Hee University, 1732 Deogyeong-daero, Giheung-gu, Yongin-si, Gyeonggi-do 17104, Republic of Korea}

\author[0000-0001-5342-8906]{Seong-Kook Lee}
\affiliation{Center for the Exploration of the Origin of the Universe (CEOU), Building 45, Seoul National University, 1 Gwanak-ro, Gwanak-gu, Seoul 08826, Republic of Korea}
\affiliation{Astronomy Program, FPRD, Department of Physics \& Astronomy, Seoul National University, 1 Gwanak-ro, Gwanak-gu, Seoul 08826, Republic of Korea}

\author{Won-Kee Park}
\affiliation{Korea Astronomy and Space Science Institute, Daejeon 34055, Republic of Korea}

\author{Yongmin Yoon}
\affiliation{Center for the Exploration of the Origin of the Universe (CEOU), Building 45, Seoul National University, 1 Gwanak-ro, Gwanak-gu, Seoul 08826, Republic of Korea}
\affiliation{Astronomy Program, FPRD, Department of Physics \& Astronomy, Seoul National University, 1 Gwanak-ro, Gwanak-gu, Seoul 08826, Republic of Korea}

\author{Seoyeon Byeon}
\affiliation{Center for the Exploration of the Origin of the Universe (CEOU), Building 45, Seoul National University, 1 Gwanak-ro, Gwanak-gu, Seoul 08826, Republic of Korea}
\affiliation{School of Space Research, Kyung Hee University, 1732 Deogyeong-daero, Giheung-gu, Yongin-si, Gyeonggi-do 17104, Republic of Korea}

\author{Sungyong Hwang}
\affiliation{Center for the Exploration of the Origin of the Universe (CEOU), Building 45, Seoul National University, 1 Gwanak-ro, Gwanak-gu, Seoul 08826, Republic of Korea}
\affiliation{Astronomy Program, FPRD, Department of Physics \& Astronomy, Seoul National University, 1 Gwanak-ro, Gwanak-gu, Seoul 08826, Republic of Korea}

\author{Joonho Kim}
\affiliation{Center for the Exploration of the Origin of the Universe (CEOU), Building 45, Seoul National University, 1 Gwanak-ro, Gwanak-gu, Seoul 08826, Republic of Korea}
\affiliation{Astronomy Program, FPRD, Department of Physics \& Astronomy, Seoul National University, 1 Gwanak-ro, Gwanak-gu, Seoul 08826, Republic of Korea}

\author{Sophia Kim}
\affiliation{Center for the Exploration of the Origin of the Universe (CEOU), Building 45, Seoul National University, 1 Gwanak-ro, Gwanak-gu, Seoul 08826, Republic of Korea}
\affiliation{Astronomy Program, FPRD, Department of Physics \& Astronomy, Seoul National University, 1 Gwanak-ro, Gwanak-gu, Seoul 08826, Republic of Korea}

\author{Gu Lim}
\affiliation{Center for the Exploration of the Origin of the Universe (CEOU), Building 45, Seoul National University, 1 Gwanak-ro, Gwanak-gu, Seoul 08826, Republic of Korea}
\affiliation{Astronomy Program, FPRD, Department of Physics \& Astronomy, Seoul National University, 1 Gwanak-ro, Gwanak-gu, Seoul 08826, Republic of Korea}

\author{Insu Paek}
\affiliation{Center for the Exploration of the Origin of the Universe (CEOU), Building 45, Seoul National University, 1 Gwanak-ro, Gwanak-gu, Seoul 08826, Republic of Korea}
\affiliation{Astronomy Program, FPRD, Department of Physics \& Astronomy, Seoul National University, 1 Gwanak-ro, Gwanak-gu, Seoul 08826, Republic of Korea}

\author{Woojin Park}
\affiliation{Center for the Exploration of the Origin of the Universe (CEOU), Building 45, Seoul National University, 1 Gwanak-ro, Gwanak-gu, Seoul 08826, Republic of Korea}
\affiliation{School of Space Research, Kyung Hee University, 1732 Deogyeong-daero, Giheung-gu, Yongin-si, Gyeonggi-do 17104, Republic of Korea}

\author{Suhyun Shin}
\affiliation{Center for the Exploration of the Origin of the Universe (CEOU), Building 45, Seoul National University, 1 Gwanak-ro, Gwanak-gu, Seoul 08826, Republic of Korea}
\affiliation{Astronomy Program, FPRD, Department of Physics \& Astronomy, Seoul National University, 1 Gwanak-ro, Gwanak-gu, Seoul 08826, Republic of Korea}




\begin{abstract}

The faint quasars with $M_{1450}>-24$ mag are known to hold the key to the determination of the ultraviolet emissivity for the cosmic re-ionization.
But only a few have been identified so far because of the limitations on the survey data.
Here, we present the first results of the $z\sim5$ faint quasar survey with the Infrared Medium-deep Survey (IMS), which covers $\sim100$ deg$^{2}$ areas in $J$-band to the depths of $J_{\rm AB}\sim23$ mag.
To improve selection methods, the medium-band follow-up imaging has been carried out using the SED camera for QUasars in Early uNiverse (SQUEAN) on the Otto Struve 2.1 m Telescope.
The optical spectra of the candidates were obtained with 8-m class telescopes.
We newly discovered 10 quasars with $-25<M_{1450}<-23$ at $z\sim5$, among which three have been missed in a previous survey using the same optical data over the same area, implying the necessity for improvements in high redshift faint quasars selection.
We derived photometric redshifts from the medium-band data, and find that they have high accuracies of $\langle|\Delta z|/(1+z)\rangle=0.016$.
The medium-band-based approach allows us to rule out many of the interlopers that contaminate $\gtrsim20~\%$ of the broad-band-selected quasar candidates. These results suggest that the medium-band-based approach is a powerful way to identify $z\sim5$ quasars and measure their redshifts at high accuracy (1-2 \%). 
It is also a cost-effective way to understand the contribution of quasars to the cosmic re-ionization history.
\end{abstract}



\section{Introduction} \label{sec:introduction}
	
Based on wide-field surveys, half million quasars have hitherto been discovered (e.g., \citealt{Paris17}),
hundreds of them being at high redshift of $z\gtrsim5$ \citep{Fan01,Fan06,Wolf03,Richards06,Fontanot07,Willott10,Mortlock11,Ikeda12,Ikeda17,McGreer13,McGreer18,
Venemans13,Venemans15a,Venemans15b,Banados14,Banados16,Banados18,Kashikawa15,Kim15,Wu15,Jun15,Jiang16,Matsuoka16,Jeon17,Yang16,Yang17,Wang16,Reed17}.
With the identification of high redshift quasars, we are now broadening our horizon of knowledge deep into the
very early universe, especially on the cosmic re-ionization epoch.
 
Recent results from the Planck collaboration suggest an instantaneous re-ionization of the intergalactic medium (IGM) at $z\sim8.8$ \citep{Planck16}, which is complete by $z\sim5$.
At $z\sim2$, we know that active galactic nuclei (AGNs) are the main IGM ionizing sources (e.g., \citealt{Haardt12}),
but at higher redshifts, stellar light from low-mass star-forming galaxies has been suggested to be the main re-ionization source
\citep{Fontanot12,Fontanot14,Robertson13,Robertson15,Japelj17,Hassan18}.
However, such a scenario has met difficulties: it requires an exceptionally large escape fraction of Lyman continuum photons
($>20$ \% of opposed to a few \% for Lyman break galaxies at $z\sim3$; \citealt{Fontanot12,Japelj17,Matthee17,Grazian17})
and/or a very steep faint end slope for the galaxy luminosity function (LF; \citealt{Bouwens17,Japelj17}).
Alternatively, \cite{Giallongo15} and \cite{Madau15} suggest that AGNs are the main IGM ionizing sources at $4<z<6.5$.
However, at $z\sim6$, results are emerging that the contribution of faint quasars to the IGM ionization is not significant (e.g., \citealt{Kim15,Onoue17}).
At $z\sim5$, it is not yet clear whether quasars or galaxies produce more ultraviolet (UV) ionizing photons.
The derivation of the LF by \cite{Giallongo15} relies on the interpolation between a photometric redshift sample of very faint quasar candidates ($M_{1450}>-22$ mag) and spectroscopically identified luminous quasars ($M_{1450}<-26$ mag).
With their LF, the major contributor of the UV luminosity density is quasars with $M_{1450}\sim-23.5$ mag.

To date, various groups have performed surveys for $z\sim5$ quasars with optical and/or infrared data
\citep{Ikeda12,Ikeda17,McGreer13,McGreer18,Jeon16,Jeon17,Yang16,Yang17}.
While most of the spectroscopically identified $z\sim5$ quasars are bright with $M_{1450}<-24$ mag,
the most recent study of \cite{McGreer18} (hereafter \citetalias{McGreer18}) focused on the dearth of quasars
at $M_{1450}\sim-23$ mag.
They found 104 candidates in the Canada-France-Hawaii Telescope Legacy Survey (CFHTLS) stacked images \citep{Gwyn12} by using the broad-band color selection method
and/or the likelihood method, and 8 of them are spectroscopically identified as faint quasars ($M_{1450}>-24$ mag)
at $4.7<z<5.4$.
The faint end of the quasar luminosity function (QLF) derived from these quasars shows a lower number density than the result from
\cite{Giallongo15} by an order of magnitude, implying low ionizing emissivity of $z\sim5$ AGNs and their minor contribution to the cosmic re-ionization.
Recent X-ray studies also suggested that the QLF of \cite{Giallongo15} could be overestimated and high redshift AGNs might
not be main contributors to the cosmic re-ionization \citep{Ricci17,Parsa18}.
At the faint end, however, the QLFs from the X-ray AGNs are still higher than that from the UV/optical survey by \citetalias{McGreer18}.
The selection methods of \citetalias{McGreer18} (both optical color selection and a likelihood method) might miss quasars, or conversely, their candidates could be contaminated by brown dwarfs or galaxies with peculiar colors,
considering the lack of near-infrared (NIR) data and the low spectral resolution for using the likelihood method.

Recently, we performed a NIR imaging survey named the Infrared Medium-deep Survey (IMS; M. Im et al., in prep), where NIR imaging data were obtained by
United Kingdom Infrared Telescope (UKIRT) at Hawaii.
The data reaches $5\sigma$ depths of $J\sim23$ mag, over 100 deg$^{2}$ areas in the sky, which overlap with the ancillary optical data from CFHTLS of which $5\sigma$ depths reach $\gtrsim25$ mag in $u'g'r'i'z'$-bands.
The combination of these optical and NIR data enables us to sample quasars as faint as $M_{1450}\sim-23$ mag at $z\sim5$.

In addition to this, we developed the SED Camera for Quasars in EArly uNiverse (SQUEAN; \citealt{Kim16,Choi15}),
as an upgraded instrument of the Camera for Quasars in EArly uNiverse (CQUEAN; \citealt{Park12,Kim11,Lim13}),
on the 2.1 m Otto Struve Telescope of McDonald Observatory.
This new instrument works with 20 filters consisting of broad-band filters (e.g., $griz$) and 50 nm medium bandwidth filters of which the central wavelengths range 675 to 1025 nm ($m675$-$m1025$\footnote{The medium-band filters are named as $m$ (initial of the medium-band) $+$ the central wavelength of the filter in nm.}).
Through observations of bright quasars at $z\sim5$, \cite{Jeon16} verified its effectiveness on distinguishing
high redshift quasars ($4.7<z<6.0$) from brown dwarfs, which are regarded as the main contaminator on high redshift quasar selection.
Furthermore, the redshift determination through the photometric redshift ($\zphot$) derived from broad- and medium-band
data shows an accuracy of 1-2 \% when compared to the spectroscopic redshift ($\zspec$).
Besides, the other surveys with medium-band observations such as COMBO-17 \citep{Wolf03}, ALHAMBRA \citep{Moles08,Matute12},
and NEWFIRM Medium-band Survey \citep{vanDokkum09} also obtained the redshifts of quasars or galaxies at $1\lesssim z \lesssim 4$ successfully with few percent uncertainties.
In addition, \cite{Matute13} discovered a faint quasar with $M_{1450}=-24.07$ mag at $z=5.41$ from the $\sim1$ deg$^{2}$ area of ALHAMBRA survey by adopting a spectral energy distribution (SED) fitting method \citep{Matute12}.
These results testify the effectiveness of using medium-band observations for the redshift determination of high redshift quasars.

Based on the optical data of CFHTLS and the NIR data of IMS, we are now performing a $z\sim5$ quasar survey with a
medium-band-based approach to increase the number of the faint quasar sample at $z\sim5$ and better determine their number density.
In this paper, we present the initial results of the $z\sim5$ quasar survey with the medium-band observations,
reporting newly discovered ten quasars at $z\sim5$ which are in the magnitude range of $-25<M_{1450}$ (mag) $<-23$.
We describe the data we used and the quasar selection method with broad-band color criteria in Section \ref{sec:inisel},
while the medium-band-based selection method with imaging follow-up with SQUEAN is described in Section \ref{sec:mbsel}.
In Section \ref{sec:spec}, the spectroscopy data we used are characterized, consisting of our spectroscopic observations and supplemental samples from literature.
We present our main results in Section \ref{sec:results}; the newly discovered quasars at $z\sim5$ and the effectiveness of the medium-band observations for finding faint quasars at $z\sim5$ and measuring their redshift accurately.
Finally, we present the implication of the newly discovered quasars to the faint-end slope of the QLF at $z\sim5$ in Section \ref{sec:implication}.
Through the paper, we adopt the cosmological parameters of $\Omega_{m}=0.3$, $\Omega_{\Lambda}=0.7$, and $H_{0}=70$ km s$^{-1}$ Mpc$^{-1}$, which are supported by previous observations (e.g., \citealt{Im97}).
All magnitudes in this paper are given in the AB system.
Note that Vega-based $J$-band magnitudes from IMS were converted to the AB system by following \cite{Hewett06}.

\section{INITIAL SAMPLE SELECTION} \label{sec:inisel}

\subsection{CFHTLS and IMS Imaging Data\label{sec:image}}

\begin{figure*}
\centering
\epsscale{0.95}
\plotone{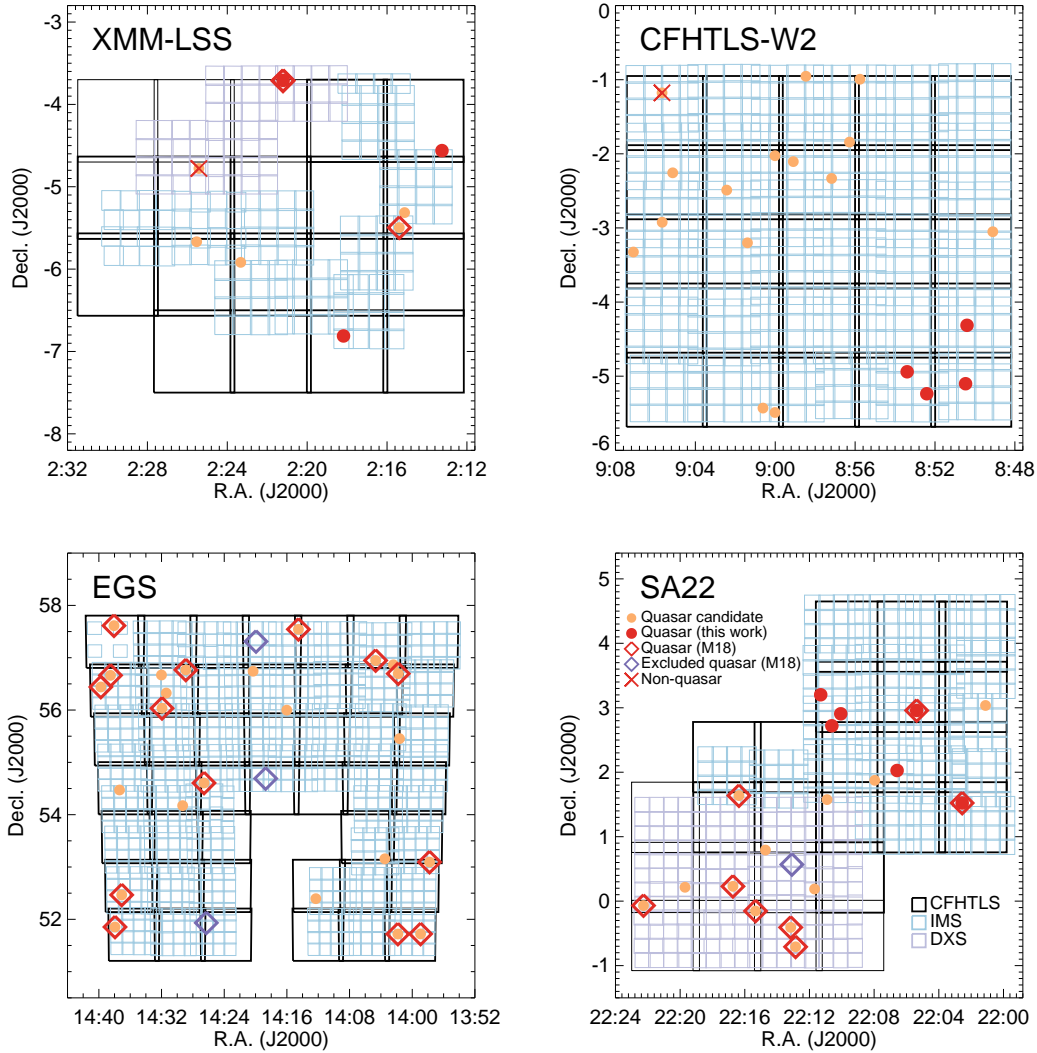}
\caption{
The coverage layout of the high-redshift quasar survey with IMS.
The four panels show the different extragalactic fields; XMM-LSS, CFHTLS-W2, EGS, and SA22.
The black squares represent the tiles of CFHTLS ($1^{\circ}\times1^{\circ}$ for each),
the blue and purple squares are the tiles of IMS and DXS ($13\farcm65\times13\farcm65$ for each), respectively.
The total survey areas of CFHTLS, IMS, and DXS in this figure are 103, 73, and 12 deg$^{2}$, respectively.
The orange filled circles represent our $z\sim5$ quasar candidates selected by broad-band color criteria,
while the spectroscopically identified quasars are additionally marked with the red filled circles (this work) and the red open diamonds (\citetalias{McGreer18}).
Note that some of the spectroscopically identified quasars with $i<23$ mag (\citetalias{McGreer18}), which are located in our survey area but excluded by our selection due to their broad-band colors, are shown as the purple open diamonds, for easy distinction.
The red crosses show the candidates spectroscopically identified as non-quasar objects.
\label{fig:field}}
\end{figure*}

Here, we describe the imaging data from which quasar candidates are selected based on the broad-band colors.
This selection is the initial step of the high redshift quasar selection, which will be refined later through medium-band imaging follow-up observation (Section \ref{sec:mbsel})
The sample selection was first carried out on the optical data from the CFHTLS Wide Survey \citep{Hudelot12} 
and the NIR data from the IMS (M. Im et al., in prep) and the Deep eXtragalactic Survey (DXS; \citealt{Lawrence07}).
There are four extragalactic fields covered by these surveys; XMM-Large Scale Structure survey region (XMM-LSS), CFHTLS Wide survey second region (CFHTLS-W2), Extended Groth Strip (EGS), and Small Selected Area 22h (SA22).
Figure \ref{fig:field} shows the positions and layouts of tiles in CFHTLS (black squares), IMS (blue squares), 
and DXS (purple squares).
Hereafter, for convenience, we call the combination of NIR data from IMS and DXS as ``IMS".

For CFHTLS, we used stacked images from the TERAPIX processing pipeline (see \citealt{Hudelot12} and the T0007 documentation file\footnote{http://terapix.iap.fr/cplt/T0007/doc/T0007-doc.html \label{test}}),
which are given for each CFHTLS tile in each CFHTLS field.
Note that ``CFHTLS tile" here denotes the $1^{\circ}\times1^{\circ}$ area named from the position of each MegaCam field of view of the Wide survey (e.g., W1+0+0),
while ``CFHTLS field" indicates the four extragalactic fields of the Wide survey (e.g., W1, W2, W3, and W4).
The zero-point ($zp$) of each tile was re-estimated by comparing the point sources in CFHTLS with those in Sloan Digital Sky Survey Data Release 12 (SDSS DR12).
Through the SQL service of SDSS, we selected point sources, 
classified as star-like sources, within the appropriate magnitude range of
$17<r<18.5$, considering the saturation level of CFHTLS and the photometric accuracy (magnitude errors $<0.1$ mag) of SDSS data in all the bands.
For the position matched sources with reliable photometry (i.e. spatially isolated point sources without saturation), 
we compared their auto magnitudes (MAG\_AUTO in \texttt{SExtractor}; \citealt{Bertin96}) of them from CFHTLS
with their PSF magnitudes from SDSS, and determined a reliable $zp$ for each tile.
In this process, we converted the auto magnitudes in optical bands ($u'g'r'i'z'$) to SDSS photometric systems ($ugriz$),
following the transformations from MegaCam to SDSS\footnote{http://www.cadc-ccda.hia-iha.nrc-cnrc.gc.ca/en/megapipe/docs/filtold.html}.
For the tiles, which do not overlap with the SDSS area, we used the overlapped stars in adjacent CFHTLS fields.
The average and standard deviation of the $zp$ value offsets in $u$, $g$, $r$, $i$, and $z$-bands are $0.14\pm0.04$, $-0.06\pm0.02$, $-0.05\pm0.02$, $-0.06\pm0.02$, and $-0.09\pm0.03$ mag, respectively.

On the other hand, for IMS, we stacked the images of each detector covering the area of $13\farcm65\times13\farcm65$ instead of
stacking the images of each IMS tile covering $0.75\times0.75$ deg$^2$ area, 
in order to determine reliable $zp$ for each image.
The $zp$ of each stacked image was scaled to 28.0 in Vega system by comparing 
the $J$-band auto-magnitudes of point sources in IMS and those from the 2MASS catalog \citep{Skrutskie06}.
The average $5\sigma$ point source detection limits of the optical/NIR images are $u=26.1$, $g=26.4$, $r=25.9$, $i=25.6$, $z=24.6$, and $J=22.9$\footnote{Unlike the homogeneous optical data, the $J$-band data including IMS and DXS is inhomogeneous. The average depths of 4 extragalactic fields of IMS (XMM-LSS, CFHTLS-W2, EGS, and SA22) are 23.2, 22.7, 22.7, and 23.2 mag, respectively, and those of DXS (XMM-LSS and SA22) are 23.7 and 23.9 mag, respectively. } mag, 
enabling us to select $z\sim5$ quasars with $i\lesssim23$ mag or those as faint as $M_{1450}\lesssim-23$ mag.
For photometry, we detected sources in the $i'$-band images
and estimated fluxes in each band within 2$\times$FWHM$_{i'}$ diameters,
using the dual-image mode of SExtractor software,
with DETECT\_THRESH of 1.3 and DETECT\_MINAREA of 9, corresponding to a $\sim4\sigma$ detection limit.
By applying aperture correction factors derived from bright stars in each filter image,
we converted the aperture magnitudes to total magnitudes.
Note that the total magnitudes were also converted to the SDSS photometric system.

\begin{figure*}
\centering
\epsscale{0.95}
\plotone{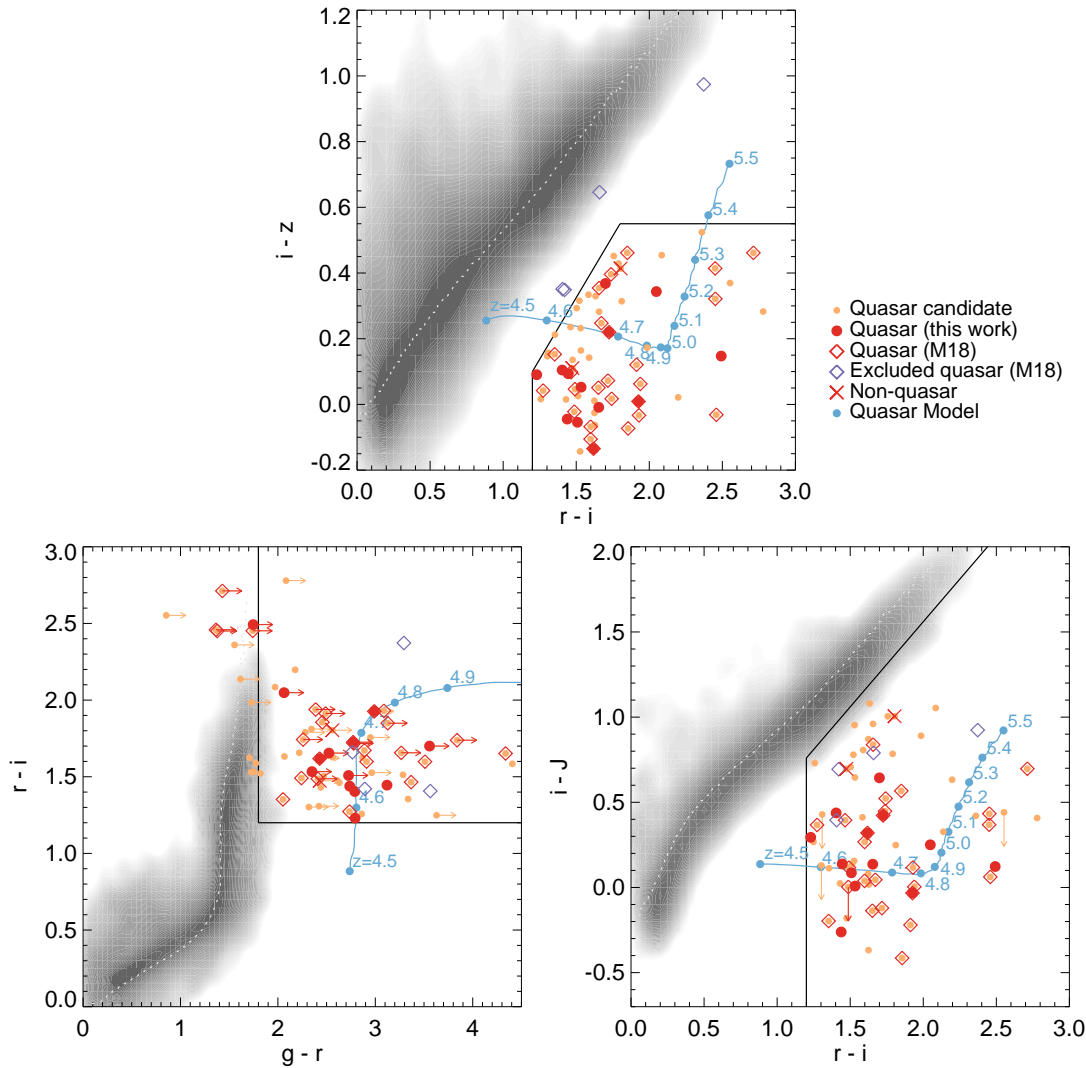}
\caption{
The color-color diagrams for quasar selection.
The gray contours represent the point sources from one of the tiles of CFHTLS (and IMS),
while the dotted lines on the contours indicate the stellar loci of \cite{Covey07}.
The black solid lines indicate our selection criteria, and the blue filled circles with lines show the redshift evolution of our quasar SED model described in Section \ref{sec:modelQ} on the color-color spaces.
The other symbols of candidates, quasars and non-quasars are same as Figure \ref{fig:field}.
Note that the arrows indicate the upper/lower limit of colors.
\label{fig:ccd}}
\end{figure*}

Although we adjusted the $zp$ values of the optical/NIR images with point sources in the SDSS/2MASS catalogs, respectively,
there are small inconsistencies of stellar loci on the order of $\lesssim0.1$ mag on color-color diagrams from tile to tile.
Compared to the stellar libraries of \cite{Pickles98}, these offsets were already reported by the TERAPIX team as one can see in their color-color diagrams$^{\ref{test}}$.
Since the color offset can affect the quasar candidates selection substantially,
we calculated the color offsets of stellar loci in each CFHTLS tile to correct the inconsistencies and improve the color selection for quasar candidates (see details in Appendix \ref{sec:coffset}).
Note that the color offsets are not adjusted for the apparent magnitudes of the quasars in this paper, but are used only for the color selection of quasar candidates in Section \ref{sec:bbcolsel}.

For the Galactic extinction correction, we used the extinction map of \cite{Schlafly11} 
with the  \cite{Cardelli89} law assuming $R_{V}=3.1$.
To account for the pixel-to-pixel correlation from the image-combining process, 
we scaled magnitude errors accordingly,
using the noise properties ($\sigma_{N}$) of an effective aperture size $N$
in each image \citep{Gawiser06,Jeon10,Kim15}.

\subsection{Broad-band Color Selection} \label{sec:bbcolsel}

\begin{deluxetable*}{lcccccccc}
\tablecaption{Broad-band Photometry of Spectroscopically Observed Quasar Candidates \label{tbl:broad}}
\tablewidth{0pt}
\tabletypesize{\scriptsize}
\tablehead{
\colhead{ID} & \colhead{R.A.} & \colhead{Decl.} & \colhead{$u$} & \colhead{$g$} & \colhead{$r$} & \colhead{$i$} & \colhead{$z$} & \colhead{$J$} \\
\colhead{} & \colhead{(J2000)} & \colhead{(J2000)} & \colhead{(mag)} & \colhead{(mag)} & \colhead{(mag)} & \colhead{(mag)} & \colhead{(mag)} & \colhead{(mag)}
}
\startdata
\multicolumn{9}{c}{Spectroscopically identified quasars}\\
IMS J021315$-$043341$^{\dagger}$$^{\ddagger}$ & 02:13:15.00 & $-$04:33:40.5 & $>26.70$  & $>26.59$  & $24.01\pm  0.09$  & $22.35\pm  0.02$  & $22.28\pm  0.06$  & $22.41\pm  0.12$  \\
IMS J021523$-$052946 & 02:15:23.29 & $-$05:29:45.9 & $>26.46$  & $>26.57$  & $22.71\pm  0.02$  & $20.99\pm  0.01$  & $20.53\pm  0.01$  & $20.75\pm  0.02$  \\
IMS J021811$-$064843$^{\dagger}$$^{\ddagger}$ & 02:18:10.80 & $-$06:48:42.6 & $>26.73$  & $25.70\pm  0.13$  & $22.87\pm  0.02$  & $21.46\pm  0.01$  & $21.20\pm  0.04$  & $21.14\pm  0.04$  \\
IMS J022112$-$034232$^{\dagger}$ & 02:21:12.32 & $-$03:42:31.8 & $>26.80$  & $26.44\pm  0.32$  & $23.41\pm  0.09$  & $21.48\pm  0.01$  & $21.38\pm  0.03$  & $21.72\pm  0.10$  \\
IMS J022113$-$034252 & 02:21:12.62 & $-$03:42:52.3 & $>26.80$  & $24.57\pm  0.06$  & $21.02\pm  0.01$  & $19.41\pm  0.00$  & $19.43\pm  0.01$  & $19.58\pm  0.03$  \\
IMS J085024$-$041850$^{\dagger}$$^{\ddagger}$ & 08:50:23.81 & $-$04:18:49.6 & $>26.31$  & $26.07\pm  0.13$  & $23.31\pm  0.03$  & $21.90\pm  0.01$  & $21.89\pm  0.04$  & $22.29\pm  0.17$  \\
IMS J085028$-$050607$^{\dagger}$$^{\ddagger}$ & 08:50:28.16 & $-$05:06:06.9 & $>26.21$  & $>26.75$  & $24.67\pm  0.14$  & $22.66\pm  0.02$  & $22.34\pm  0.06$  & $22.52\pm  0.28$  \\
IMS J085225$-$051413$^{\dagger}$$^{\ddagger}$ & 08:52:24.73 & $-$05:14:13.4 & $>26.59$  & $>26.50$  & $24.15\pm  0.09$  & $22.64\pm  0.02$  & $22.61\pm  0.06$  & $22.75\pm  0.26$  \\
IMS J085324$-$045626$^{\dagger}$$^{\ddagger}$ & 08:53:23.68 & $-$04:56:25.6 & $>26.53$  & $>26.48$  & $23.75\pm  0.08$  & $22.27\pm  0.02$  & $22.35\pm  0.06$  & $22.30\pm  0.11$  \\
IMS J135747$+$530543 & 13:57:47.34 & $+$53:05:42.6 & $>26.31$  & $>26.19$  & $23.09\pm  0.04$  & $21.21\pm  0.01$  & $20.72\pm  0.02$  & $20.83\pm  0.03$  \\
IMS J135856$+$514317 & 13:58:55.96 & $+$51:43:17.0 & $>26.66$  & $26.37\pm  0.19$  & $21.99\pm  0.02$  & $20.40\pm  0.00$  & $20.29\pm  0.01$  & $20.77\pm  0.03$  \\
IMS J140147$+$564145 & 14:01:46.97 & $+$56:41:44.8 & $>26.50$  & $26.63\pm  0.20$  & $23.54\pm  0.05$  & $21.67\pm  0.01$  & $21.60\pm  0.04$  & $21.71\pm  0.07$  \\
IMS J140150$+$514310 & 14:01:49.96 & $+$51:43:10.4 & $>26.72$  & $>26.73$  & $25.33\pm  0.13$  & $22.93\pm  0.03$  & $22.90\pm  0.07$  & $23.10\pm  0.22$  \\
IMS J140440$+$565651 & 14:04:40.29 & $+$56:56:50.7 & $>26.89$  & $24.62\pm  0.05$  & $22.36\pm  0.01$  & $20.94\pm  0.00$  & $20.86\pm  0.01$  & $20.99\pm  0.03$  \\
IMS J141432$+$573234 & 14:14:31.56 & $+$57:32:34.4 & $>26.85$  & $>26.67$  & $23.45\pm  0.05$  & $21.81\pm  0.01$  & $21.42\pm  0.03$  & $21.18\pm  0.11$  \\
IMS J142635$+$543623 & 14:26:34.86 & $+$54:36:22.7 & $>26.86$  & $24.41\pm  0.04$  & $21.51\pm  0.01$  & $19.92\pm  0.00$  & $19.89\pm  0.01$  & $19.78\pm  0.02$  \\
IMS J142854$+$564602 & 14:28:53.85 & $+$56:46:02.0 & $>27.13$  & $26.64\pm  0.30$  & $23.75\pm  0.05$  & $22.07\pm  0.01$  & $21.75\pm  0.04$  & $22.26\pm  0.15$  \\
IMS J143156$+$560201 & 14:31:56.36 & $+$56:02:00.9 & $>27.16$  & $24.76\pm  0.06$  & $22.01\pm  0.01$  & $20.74\pm  0.00$  & $20.62\pm  0.02$  & $20.60\pm  0.04$  \\
IMS J143705$+$522801 & 14:37:05.17 & $+$52:28:00.8 & $>26.75$  & $>26.69$  & $24.20\pm  0.08$  & $22.30\pm  0.02$  & $22.10\pm  0.04$  & $22.75\pm  0.16$  \\
IMS J143757$+$515115 & 14:37:56.54 & $+$51:51:15.1 & $>26.90$  & $>26.68$  & $24.91\pm  0.07$  & $22.43\pm  0.03$  & $22.07\pm  0.05$  & $22.20\pm  0.12$  \\
IMS J143804$+$573646 & 14:38:04.05 & $+$57:36:46.4 & $>26.56$  & $>26.49$  & $24.08\pm  0.07$  & $22.60\pm  0.02$  & $22.61\pm  0.04$  & $>22.86$  \\
IMS J143831$+$563946 & 14:38:30.83 & $+$56:39:46.4 & $>26.49$  & $25.01\pm  0.09$  & $22.98\pm  0.03$  & $21.56\pm  0.01$  & $21.39\pm  0.04$  & $22.03\pm  0.15$  \\
IMS J143945$+$562627 & 14:39:44.88 & $+$56:26:26.6 & $>26.60$  & $>26.72$  & $24.48\pm  0.09$  & $22.67\pm  0.03$  & $22.63\pm  0.10$  & $22.42\pm  0.23$  \\
IMS J220233$+$013120$^{\dagger}$ & 22:02:33.20 & $+$01:31:20.3 & $>26.53$  & $>26.84$  & $24.02\pm  0.07$  & $22.26\pm  0.03$  & $21.95\pm  0.04$  & $22.02\pm  0.08$  \\
IMS J220522$+$025730$^{\dagger}$ & 22:05:22.15 & $+$02:57:30.0 & $>26.47$  & $25.80\pm  0.13$  & $23.32\pm  0.05$  & $21.66\pm  0.01$  & $21.68\pm  0.05$  & $21.43\pm  0.08$  \\
IMS J220635$+$020136$^{\dagger}$$^{\ddagger}$ & 22:06:34.81 & $+$02:01:36.3 & $>26.06$  & $>26.31$  & $24.58\pm  0.04$  & $22.08\pm  0.02$  & $21.81\pm  0.05$  & $22.08\pm  0.10$  \\
IMS J221004$+$025424$^{\dagger}$$^{\ddagger}$ & 22:10:03.90 & $+$02:54:24.4 & $>26.65$  & $26.37\pm  0.15$  & $23.56\pm  0.06$  & $22.36\pm  0.01$  & $22.30\pm  0.05$  & $22.19\pm  0.08$  \\
IMS J221037$+$024314$^{\dagger}$$^{\ddagger}$ & 22:10:36.99 & $+$02:43:13.7 & $>26.60$  & $>26.71$  & $23.13\pm  0.04$  & $21.46\pm  0.00$  & $21.12\pm  0.02$  & $20.95\pm  0.03$  \\
IMS J221118$+$031207$^{\dagger}$$^{\ddagger}$ & 22:11:18.37 & $+$03:12:07.4 & $>26.71$  & $25.91\pm  0.09$  & $22.77\pm  0.02$  & $21.36\pm  0.00$  & $21.29\pm  0.02$  & $21.35\pm  0.04$  \\
IMS J221251$-$004231 & 22:12:51.49 & $-$00:42:30.7 & $>26.78$  & $24.33\pm  0.04$  & $21.77\pm  0.01$  & $19.91\pm  0.00$  & $19.89\pm  0.00$  & $20.48\pm  0.04$  \\
IMS J221310$-$002428 & 22:13:09.67 & $-$00:24:28.1 & $>26.88$  & $>27.08$  & $24.59\pm  0.15$  & $22.65\pm  0.02$  & $22.49\pm  0.05$  & $22.80\pm  0.21$  \\
IMS J221520$-$000908 & 22:15:20.22 & $-$00:09:08.4 & $>26.54$  & $>26.52$  & $25.05\pm  0.09$  & $22.19\pm  0.03$  & $21.77\pm  0.05$  & $21.67\pm  0.07$  \\
IMS J221622$+$013815 & 22:16:21.85 & $+$01:38:14.7 & $>26.42$  & $>26.73$  & $25.32\pm  0.11$  & $22.85\pm  0.03$  & $22.35\pm  0.05$  & $22.63\pm  0.12$  \\
IMS J221644$+$001348 & 22:16:44.02 & $+$00:13:48.2 & $>26.40$  & $25.48\pm  0.16$  & $22.07\pm  0.01$  & $20.54\pm  0.00$  & $20.34\pm  0.01$  & $20.26\pm  0.04$  \\
IMS J222216$-$000406 & 22:22:16.02 & $-$00:04:05.7 & $>26.28$  & $>26.60$  & $23.78\pm  0.06$  & $22.02\pm  0.01$  & $21.83\pm  0.04$  & $22.26\pm  0.11$  \\
\hline
\multicolumn{9}{c}{Spectroscopically identified non-quasars}\\
IMS J022525$-$044642 & 02:25:25.18 & $-$04:46:41.5 & $>26.85$  & $>26.70$  & $24.27\pm  0.12$  & $22.85\pm  0.04$  & $22.64\pm  0.10$  & $22.37\pm  0.17$  \\
IMS J090540$-$011038 & 09:05:40.10 & $-$01:10:38.4 & $>26.80$  & $>26.58$  & $24.02\pm  0.07$  & $22.20\pm  0.02$  & $21.68\pm  0.04$  & $21.34\pm  0.10$  \\
\enddata
\tablecomments{
All magnitudes are given in AB system, and their errors are scaled with $\sigma_{N}$, the noise properties of an effective aperture size $N$ (see details in Section \ref{sec:image}).
Note that all the magnitudes are not corrected with $C^{\rm offset}_{k}$ (also see Section \ref{sec:image} and Appendix \ref{sec:coffset}).
}
\tablenotetext{\dagger}{These quasars are spectroscopically observed in this work.}
\tablenotetext{\ddagger}{These quasars are newly discovered in this work, while others are discovered by previous studies (\citealt{McGreer13}; \citetalias{McGreer18}; \citealt{Ikeda17}).}
\end{deluxetable*}

The broad-band color selection follows the criteria of \cite{McGreer13},
where they defined the color selection by simulating the color tracks using low redshift SDSS quasar spectra
that are redshifted to $z\sim5$.
Considering the deeper depths of CFHTLS and IMS, we made a minor change to the $i$-magnitude limit.
The following shows the selection criteria that we used:

\begin{enumerate}
\item $i < 23$,
\item S/N ($u) < 2.5$,
\item $g-r>1.8$ or S/N ($g)<3.0$,
\item $r-i>1.2$,
\item $i-z<0.625~((r-i)-1.0)$,
\item $i-z<0.55$,
\item $i-J<( (r-i) -1.0) + 0.56$,
\end{enumerate}

\begin{figure*}
\centering
\epsscale{1}
\plotone{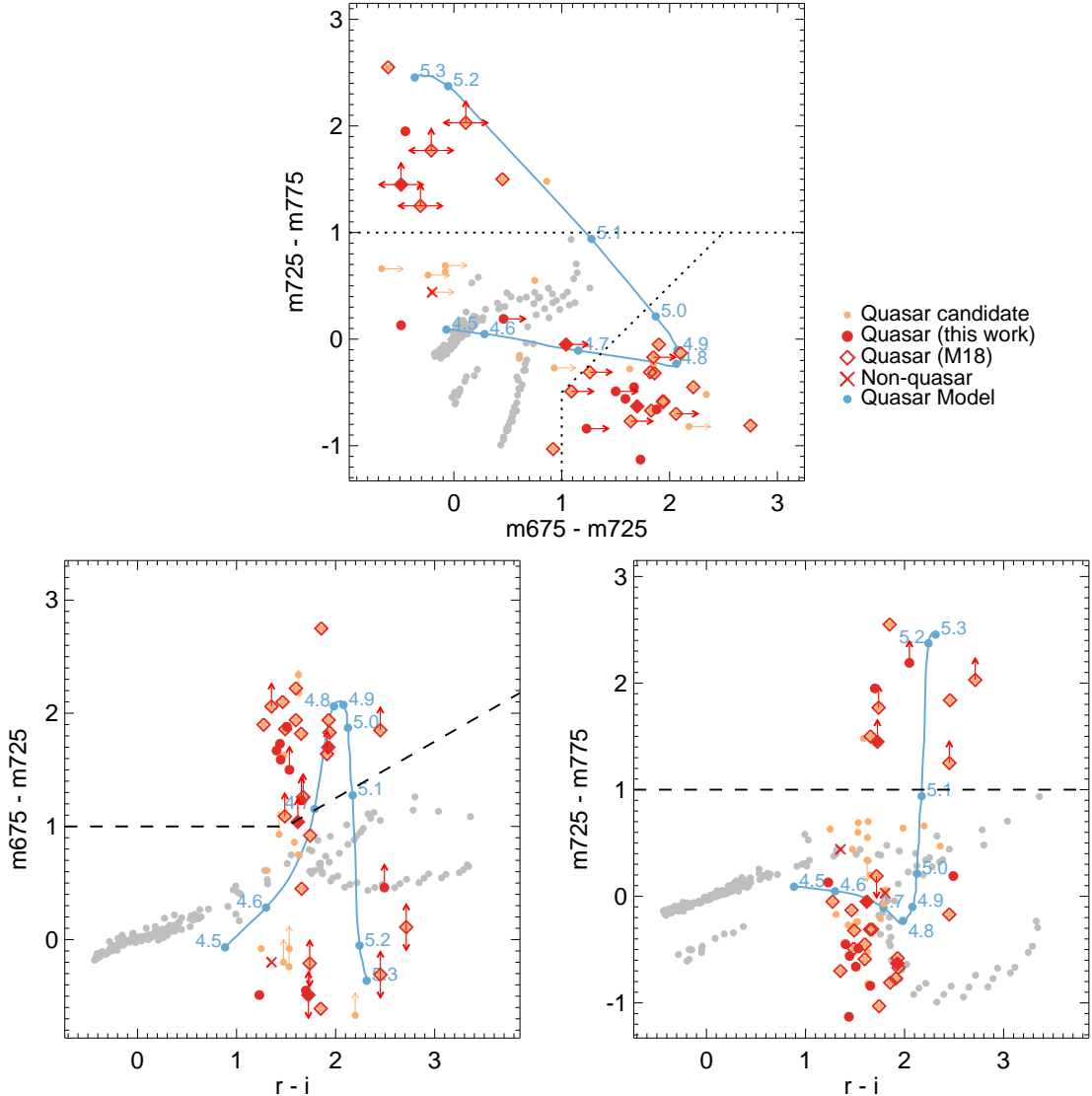}
\caption{
The medium-band colors of quasars.
The symbols of candidates, quasars, non-quasars, and quasar model are same as Figure \ref{fig:ccd}.
The gray filled circles represent the colors of typical stars from stellar templates of main sequence \citep{Gunn83} and dwarf \citep{Burrows06}.
The dotted lines are the medium-band selection criteria provided by \cite{Jeon16},
while the dashed lines are the additional criteria presented in this work.
\label{fig:ccd2}}
\end{figure*}

\noindent where the S/N values are directly estimated from the fluxes and flux errors in the aperture mentioned above.
The candidates satisfying the criteria were visually inspected to exclude spurious objects 
such as cross-talks, diffraction spikes, etc., resulting in 70 $z\sim5$ quasar candidates.
The positions of the candidates (orange circles) are plotted on the layouts in Figure \ref{fig:field}.
Figure \ref{fig:ccd} shows the color-color diagrams 
($g-r$ vs. $r-i$, $r-i$ vs. $i-z$, and $r-i$ vs. $i-J$) of objects in the multi-band catalog and the broad-band color selection criteria (the black solid lines).
The broad-band photometry of our candidates are listed in Table \ref{tbl:broad}.
In this paper, we only include the candidates, which are spectroscopically observed in this work or previous works (e.g., \citetalias{McGreer18}), and also observed in medium-bands, instead of the full sample of our candidates (see details of the spectroscopic sample in Section \ref{sec:spec}).

\begin{deluxetable*}{ccccccccccccc}
\tabletypesize{\scriptsize}
\tablecaption{Medium-band Photometry of Spectroscopically Observed Quasar Candidates \label{tbl:medium}}
\tablewidth{0pt}
\tablehead{
\colhead{ID} & \multicolumn{12}{c}{Observing Runs, exposure times (s), and magnitudes (mag)} \\
\cline{2-13}
\colhead{} & \multicolumn{3}{c}{$m675$} & \multicolumn{3}{c}{$m725$} & \multicolumn{3}{c}{$m775$} & \multicolumn{3}{c}{$m825$}
}
\startdata
\multicolumn{13}{c}{Spectroscopically identified quasars}\\
IMS J021315$-$043341 & 17Oct & 3600 & $>23.82$  & 15Oct & 1800 & $22.59\pm  0.17$  & 15Oct & 4140 & $23.43\pm  0.80$  & 17Oct & 2700 & $22.86\pm  0.26$  \\
IMS J021523$-$052946 & 17Dec & 1800 & $>22.33$  & 17Dec & 1800 & $>22.54$  & 17Dec & 1260 & $20.77\pm  0.23$  & - & -   & -  \\
IMS J021811$-$064843 & 17Oct & 1800 & $22.92\pm  0.17$  & 16Feb & 900 & $21.25\pm  0.13$  & 16Feb & 900 & $21.70\pm  0.27$  & 17Oct & 1800 & $21.57\pm  0.07$  \\
IMS J022112$-$034232 & 17Sep & 1800 & $22.97\pm  0.54$  & 16Feb & 900 & $21.27\pm  0.08$  & 16Feb & 900 & $21.90\pm  0.28$  & 17Sep & 1800 & $21.79\pm  0.22$  \\
IMS J022113$-$034252 & 17Dec & 600 & $21.27\pm  0.17$  & 17Dec & 300 & $19.33\pm  0.04$  & 17Dec & 300 & $19.92\pm  0.06$  & - & -   & -  \\
IMS J085024$-$041850 & 17Dec & 2700 & $23.67\pm  0.73$  & 17Dec & 1800 & $21.94\pm  0.11$  & 17Dec & 1800 & $23.07\pm  0.34$  & - & -   & -  \\
IMS J085028$-$050607 & - & -   & -  & 17Apr & 3600 & $>23.97$  & 17Apr & 3600 & $21.78\pm  0.10$  & 18Apr & 3600 & $22.63\pm  0.16$  \\
IMS J085225$-$051413 & 18Apr & 3600 & $>24.03$  & 17Apr & 3600 & $22.53\pm  0.10$  & 17Apr & 3600 & $23.02\pm  0.26$  & - & -   & -  \\
IMS J085324$-$045626 & 17Dec & 3240 & $24.25\pm  0.60$  & 16Feb & 1800 & $22.37\pm  0.15$  & 16Feb & 1800 & $23.03\pm  0.40$  & - & -   & -  \\
IMS J135747$+$530543 & 18Jan & 2700 & $22.47\pm  0.34$  & 17Apr & 3600 & $23.08\pm  0.20$  & 17Apr & 3600 & $20.53\pm  0.08$  & 17Dec & 900 & $21.22\pm  0.12$  \\
IMS J135856$+$514317 & 17Feb & 900 & $22.37\pm  0.12$  & 16Feb & 900 & $20.55\pm  0.05$  & 16Feb & 900 & $20.86\pm  0.09$  & - & -   & -  \\
IMS J140147$+$564145 & 17Feb & 1800 & $22.82\pm  0.19$  & 16Feb & 900 & $20.88\pm  0.04$  & 16Feb & 960 & $21.46\pm  0.08$  & - & -   & -  \\
IMS J140150$+$514310 & - & -   & -  & 17Apr & 3600 & $24.12\pm  0.35$  & 17Apr & 3600 & $22.28\pm  0.15$  & 18Apr & 3600 & $23.19\pm  0.22$  \\
IMS J140440$+$565651 & 17Dec & 3540 & $23.24\pm  0.40$  & 17Apr & 900 & $21.38\pm  0.06$  & 17Apr & 1440 & $21.70\pm  0.09$  & 16Apr & 900 & $21.61\pm  0.09$  \\
IMS J141432$+$573234 & 17Feb & 3600 & $23.20\pm  0.18$  & 16Feb & 900 & $22.75\pm  0.23$  & 16Feb & 900 & $21.25\pm  0.10$  & 17Dec & 1620 & $22.12\pm  0.20$  \\
IMS J142635$+$543623 & 17Dec & 300 & $21.86\pm  0.19$  & 17Dec & 180 & $19.64\pm  0.04$  & 17Dec & 180 & $20.09\pm  0.06$  & 17Dec & 180 & $20.06\pm  0.07$  \\
IMS J142854$+$564602 & 17Dec & 2700 & $>23.89$  & 17Apr & 4140 & $22.63\pm  0.15$  & 17Apr & 1800 & $22.94\pm  0.36$  & - & -   & -  \\
IMS J143156$+$560201 & 17Feb & 900 & $22.99\pm  0.30$  & 17Apr & 1800 & $21.09\pm  0.11$  & 17Apr & 2340 & $21.14\pm  0.20$  & 16Apr & 900 & $21.04\pm  0.06$  \\
IMS J143705$+$522801 & 17Dec & 2700 & $>23.78$  & 17Dec & 900 & $22.14\pm  0.12$  & 17Dec & 1800 & $22.91\pm  0.28$  & - & -   & -  \\
IMS J143757$+$515115 & 18Feb & 2700 & $>23.57$  & 17Apr & 3600 & $>23.88$  & 17Apr & 1800 & $22.63\pm  0.34$  & 18Feb & 1980 & $22.97\pm  0.42$  \\
IMS J143804$+$573646 & 17Dec & 2700 & $>23.70$  & 17Apr & 4140 & $22.61\pm  0.17$  & 17Apr & 5400 & $23.10\pm  0.61$  & 18Apr & 2040 & $22.86\pm  0.33$  \\
IMS J143831$+$563946 & 17Feb & 6300 & $>23.39$  & 16Jun & 900 & $21.33\pm  0.14$  & 16Jun & 900 & $22.03\pm  0.22$  & 18Apr & 1260 & $22.04\pm  0.22$  \\
IMS J143945$+$562627 & 17Dec & 2700 & $23.51\pm  0.30$  & 17Apr & 4500 & $22.59\pm  0.15$  & 17Apr & 1800 & $23.62\pm  0.97$  & - & -   & -  \\
IMS J220233$+$013120 & 16Dec & 3060 & $>23.18$  & 15Oct & 2160 & $>23.67$  & 15Oct & 2160 & $22.22\pm  0.11$  & 16Dec & 1800 & $>22.17$  \\
IMS J220522$+$025730 & 16Jul & 900 & $>22.73$  & 15Oct & 1260 & $21.69\pm  0.07$  & 15Oct & 1260 & $21.74\pm  0.09$  & 16Dec & 900 & $22.32\pm  0.31$  \\
IMS J220635$+$020136 & 16Dec & 1620 & $>22.53$  & 16Jun & 1980 & $22.07\pm  0.17$  & 16Jun & 1800 & $21.88\pm  0.18$  & 17Oct & 1800 & $22.04\pm  0.11$  \\
IMS J221004$+$025424 & 16Dec & 4200 & $22.36\pm  0.27$  & 15Oct & 1800 & $22.85\pm  0.20$  & 15Oct & 1800 & $22.72\pm  0.19$  & 16Dec & 1800 & $23.33\pm  0.56$  \\
IMS J221037$+$024314 & 16Dec & 1860 & $22.67\pm  0.49$  & 15Oct & 2520 & $23.12\pm  0.41$  & 15Oct & 1620 & $21.17\pm  0.08$  & 17Oct & 2520 & $21.36\pm  0.06$  \\
IMS J221118$+$031207 & 16Jul & 900 & $23.10\pm  0.38$  & 15Oct & 1800 & $21.51\pm  0.11$  & 15Oct & 1800 & $22.07\pm  0.20$  & 16Dec & 1440 & $21.77\pm  0.18$  \\
IMS J221251$-$004231 & 17Dec & 600 & $22.48\pm  0.37$  & 17Dec & 300 & $19.73\pm  0.06$  & 17Dec & 600 & $20.54\pm  0.12$  & - & -   & -  \\
IMS J221310$-$002428 & 17Oct & 5400 & $24.07\pm  0.37$  & 16Jun & 3600 & $22.24\pm  0.14$  & 16Jun & 3600 & $22.91\pm  0.35$  & 17Oct & 4320 & $22.94\pm  0.22$  \\
IMS J221520$-$000908 & 17Oct & 5040 & $>23.51$  & 15Oct & 1800 & $>23.40$  & 15Oct & 1800 & $21.37\pm  0.10$  & 17Oct & 2700 & $21.94\pm  0.14$  \\
IMS J221622$+$013815 & 17Oct & 5400 & $>24.57$  & 15Oct & 3780 & $22.72\pm  0.13$  & 15Oct & 3600 & $22.89\pm  0.22$  & 17Oct & 3600 & $22.89\pm  0.18$  \\
IMS J221644$+$001348 & 17Dec & 1560 & $22.37\pm  0.37$  & 17Dec & 540 & $20.27\pm  0.07$  & 17Dec & 600 & $20.40\pm  0.12$  & - & -   & -  \\
IMS J222216$-$000406 & - & -   & -  & 17Dec & 1080 & $21.65\pm  0.23$  & 18Jan & 900 & $>21.46$  & - & -   & -  \\
\hline
\multicolumn{13}{c}{Spectroscopically identified non-quasars}\\
IMS J022525$-$044642 & 17Oct & 3600 & $>23.46$  & 15Oct & 3600 & $23.66\pm  0.31$  & 15Oct & 3420 & $23.22\pm  0.30$  & - & -   & -  \\
IMS J090540$-$011038 & - & -   & -  & 17Apr & 3600 & $21.93\pm  0.20$  & 17Apr & 3600 & $21.90\pm  0.33$  & - & -   & -  \\
\enddata
\tablecomments{
All magnitudes are given in AB system, and their errors are scaled with $\sigma_{N}$.
}
\end{deluxetable*}

\section{MEDIUM-BAND SELECTION}\label{sec:mbsel}

\subsection{Medium-band Observation}\label{sec:obs_squean}

To further exclude interlopers and better determine redshifts photometrically,
we observed our candidates in medium-bands with SQUEAN from 2015 December to 2018 April.
Since the Lyman-$\alpha$ (Ly$\alpha$; 1216 $\rm\AA$) break of a $z\sim5$ quasar is expected to be located at $\lambda_{\rm obs} \sim 7300~\rm{\AA}$,
the medium-band observations were performed mainly with $m725$ and $m775$ filters.
If the two medium-band data were not enough to identify the object as a $z>5.1$ quasar (i.e. $m725-m775<1$; see Section \ref{sec:mbcolsel}),
additional imaging data in $m675$-band were also obtained.
For the spectroscopically identified quasars, if needed, observations in $m675$- and/or $m825$-bands were
also carried out to check the accuracy of the $\zphot$ from medium-band data.
For each band, we took 3-70 frames with exposure times of 1 to 3 min, which gives the total integration time of 0.05-1.75 hours per band per filter.
Note that brighter candidates ($i<22$ mag) were observed as high priority targets, when the observing condition was unstable with seeing size of $>1\farcs2$.
Among the 70 quasar candidates, 58 candidates were observed in $m725$- and $m775$-bands
and 45 of them were further observed in $m675$-band.

We reduced the medium-band data, following the procedure in \cite{Jeon16}.
After subtracting the bias and dark frames,
we divided the science frames by the normalized flat frames, which were produced from the twilight sky.
Excluding the images taken under bad weather conditions (e.g. low signals due to heavy clouds),
the science images after the reduction were combined.
We first detected the sources in the combined images with a detection threshold of $\sim2.7\sigma$
(DETECT\_THRESH of 1.2 and DETECT\_MINAREA of 5).
The $zp$ of each medium-band image was determined by fitting the stellar templates to the broad-band photometry ($riz$) of stars in each field (see details in \citealt{Jeon16}).
Note that we regarded auto-magnitudes of the stars in each medium-band as total magnitudes for the $zp$ determination.
The uncertainty in the $zp$ determination is found to be $\sim0.03$ mag, by taking the standard deviation of the $zp$ values from th stars in the same field.
For each quasar candidate, we estimated the aperture magnitude 
(size of $2\times{\rm FWHM}_{mb}$ is used, where FWHM$_{mb}$ is FWHM of point sources in each medium-band image)
with forced photometry on the target position determined in the $i$-band image.
We applied the aperture correction factor determined from the stars in each field.
Like the broad-band photometry, the Galactic extinction was corrected by following the \cite{Cardelli89} law assuming $R_{V}=3.1$ with the extinction map of \cite{Schlafly11} and also scaled the SExtractor-derived magnitude errors to account for the correlated noise in the stacked image ($\sigma_{N}$).
We gave the upper limit, which is defined as the magnitude limit for the $2.7\sigma$ detection,
to the objects with no detection or the magnitudes less than the upper limit.
The observing runs and the medium-band photometry are given in Table \ref{tbl:medium}.
As with Table \ref{tbl:broad}, the spectroscopically examined candidates are only listed.

\subsection{Medium-band Selection of $z\sim5$ Quasar Candidates} \label{sec:mbcolsel}

Figure \ref{fig:ccd2} shows the color-color diagrams for the medium-bands only (top panel for $m675-m725$ vs. $m725-m775$), and for the combinations of broad- and medium-band colors (bottom panels for $r-i$ vs. $m675-m725$ and $r-i$ vs. $m725-m775$, respectively).
The gray filled circles represent the colors of the 175 star templates covering various spectral types and luminosity classes \citep{Gunn83} and the 41 L/T dwarf star models \citep{Burrows06}. 
The other symbols are identical to those in Figure \ref{fig:ccd}.
We followed the color selection criteria with medium-bands suggested by \cite{Jeon16}:

\begin{enumerate}
\item $m675-m725>1$ and $m675-m725>m725-m775+1.5$ ($4.7<z<5.1$),
\item $m725-m775>1$ ($5.1<z<5.5$),
\end{enumerate}

\noindent which are plotted as dotted lines in Figure \ref{fig:ccd2}.
The top panel in the figure shows the above criteria at a glance.
Among 45 candidates observed in $m675$-, $m725$-, and $m775$-bands, 33 candidates satisfy the above color selection criteria.
The medium-band color criteria ($m675-m725>1$ and $m725-m775>1$) could be roughly adopted to the combination of 
broad- and medium-band colors (dashed lines).
Note that the former criterion is limited by $r-i$ color; $m675-m725> 0.5~(r-i)-0.25$.

\begin{deluxetable*}{ccccc}
\tabletypesize{\scriptsize}
\tablecaption{Spectroscopic Observations of $z\sim5$ Quasar Candidates\label{tbl:obs_spec}}
\tablewidth{0pt}
\tablehead{
\colhead{ID} & \colhead{Telescope/instrument} & \colhead{Date} & \colhead{Exposure time (s)} & \colhead{Seeing ($\arcsec$)} }
\startdata
\multicolumn{5}{c}{Spectroscopically identified quasars}\\
IMS J021315$-$043341 & Magellan/IMACS & 2016 Dec 4-5 & 4500 & 0.5-0.8 \\
IMS J021811$-$064843 & Gemini/GMOS-S  & 2016 Sep 6   & 480  & 1.0-1.1 \\
IMS J022112$-$034232 & Gemini/GMOS-S  & 2016 Sep 3   & 960  & 1.2-1.3 \\
IMS J085024$-$041840 & Gemini/GMOS-N  & 2018 May 18  & 1440 & 0.7 \\
IMS J085028$-$050607 & Gemini/GMOS-S  & 2018 Mar 20  & 3000 & 1.1 \\
IMS J085225$-$051413 & Gemini/GMOS-S  & 2018 Mar 20  & 3000 & 1.1 \\
IMS J085324$-$045626 & Magellan/IMACS & 2016 Dec 6   & 3600 & 0.6-0.9 \\
IMS J220233$+$013120 & Gemini/GMOS-S  & 2016 Sep 4-6 & 2880 & 1.1-1.3 \\
IMS J220522$+$025730 & Gemini/GMOS-S  & 2016 Sep 6   & 1440 & 1.1 \\
IMS J220635$+$020136 & Gemini/GMOS-S  & 2018 Jun 18  & 1440 & 0.8 \\
IMS J221004$+$025424 & Gemini/GMOS-S  & 2016 Sep 8   & 2880 & 0.5 \\ 
IMS J221037$+$024314 & Gemini/GMOS-S$^{a}$  & 2016 Sep 8   & 9600 & 0.8 \\ 
IMS J221118$+$031207 & Gemini/GMOS-S  & 2016 Sep 4   & 960  & 1.2-1.3 \\
\hline
\multicolumn{5}{c}{Spectroscopically identified non-quasars}\\
IMS J022525$-$044642 & Gemini/GMOS-S  & 2016 Sep 4-8 & 5760 & 1.0 \\
IMS J090540$-$011038 & Gemini/GMOS-N  & 2018 May 18  & 1440 & 0.7 \\
\enddata
\tablenotetext{a}{MOS observation with our candidate for a faint quasar at $z\sim6$ (See details in Section \ref{sec:obs_gmos})}
\end{deluxetable*}

\section{Spectroscopy Data} \label{sec:spec}

We performed spectroscopic observations of 15 candidates from the broad-band selection method,
among which 10 satisfy the medium-band selection.
The medium-band-selected candidates were spectroscopically observed prior to other candidates. Here, "other candidates" mean the objects that are outside the medium-band selection boxes but could be included considering their large magnitude uncertainties (or upper limits of flux at short wavelength).
These observations are reported below in Table \ref{tbl:obs_spec}.
Additionally, we took spectra of seven candidates from the broad-band photometry, \emph{before} we improved the photometry as described in Section \ref{sec:image}.
After improving the photometry as described in Section \ref{sec:image},
they turned out not to satisfy the broad-band quasar selection criteria and they are all found to be non-quasars from spectroscopy.
For completeness, we present these non-quasar spectra in Appendix \ref{sec:nonqso}, but we will exclude them in our analysis hereafter.
Additionally, we used published redshifts for some of the medium-band observed objects, as described in Section \ref{sec:spec_sup}.

\subsection{Gemini/GMOS Observation}\label{sec:obs_gmos}

Spectroscopic observations of 13 candidates were carried out with Gemini Multi-Object Spectrographs (GMOS; \citealt{Hook04}) on Gemini North and South 8 m Telescopes at Mauna Kea, Hawaii and Cerro Pachon, Chile, respectively,
on 2016 September 3-8 (PID: GS-2016B-Q-46), 2018 March 20 and June 18 (PID: GS-2018A-Q-220), and 2018 May 18 (PID: GN-2018A-Q-315).
The sky was almost clear with average seeings of $\sim1\farcs0$.
To ease the sky subtraction for the faint targets, the Nod \& Shuffle (N\&S) observing mode was adopted with a 1$\farcs$0 width N\&S slit.
The spectra were obtained by using the R150+\_G5326 grating which has a resolution of $R\sim 315$ at 717 nm for a slit width of $1\farcs0$, and the GG455\_G0329 or OG515\_G0330 filters to avoid the 0-th order overlap.
This set-up gives the wavelength range of 4550 or 5150 to 10300 $\rm\AA$.
In order to cover the gaps between the chips on the Hamamatsu CCD, the central wavelengths were set to 7100 and 7250 $\rm{\AA}$.
This setting allows the detection of the redshifted Ly$\alpha$ break, which is expected to be located at $\sim7200~\rm{\AA}$ for $z\sim5$ quasars.
For observing run in the 2018A semester, we set the central wavelengths to 4300 and 4600 $\rm{\AA}$ for the Gemini-South in order to avoid the bad columns on the CCD, and 6350 and 6650 $\rm{\AA}$ for the Gemini-North.
Note that we adopted a $4\times4$ binning in spatial/spectral pixels to maximize the S/N.

For one target, IMS J221046$+$024313, we obtained its spectrum
through the MOS observing mode of GMOS-S (PID: GS-2016B-Q-11) during which we observed other targets of interest for another program.
For the MOS observation with the N\&S mode, we used the same R150+\_G5326 grating with RG610\_G0331 filter, and the central wavelengths were set to 8900 and 9000 $\rm{\AA}$.
To increase the S/N, the spectrum was also binned with $4\times4$.

For data reduction, the spectra were processed by using the Gemini IRAF package.
After the bias subtraction and flat-fielding, sky lines were subtracted with the shuffled spectra.
The wavelength calibration was done with CuAr arc lines,
and the flux calibration was done with standard stars (LTT7379, CD329927, and Wolf1346).
For IMS J221036+024313 with the MOS observation, the wavelength calibration preceded the sky subtraction
due to the alignments of sky lines in the spatial direction.
The aperture size for the spectral extraction was set at $1\farcs0$ in diameter for all cases.
Note that the overall flux scale of each spectrum was adjusted using the $i$-magnitude of each target.
In order to increase the S/N, we binned the spectra along the spectral direction by a factor of 2-5 (pixels)
by using the inverse-variance weighting method (e.g., \citealt{Kim18}).
This binning gives the spectral resolution of $\sim300$.

\subsection{Magellan/IMACS Observation}\label{sec:obs_imacs}

The optical spectra of the other two candidates were obtained by Inamori-Magellan Areal Camera and Spectrograph (IMACS; \citealt{Dressler11}) on the Magellan Baade 6.5 m Telescope in Las Campanas Observatory, Chile, on 2016 December 3-5.
Unlike the Gemini observations, the Magellan spectra were obtained with a standard long-slit mode (not N\&S).
We used the f/4 camera of IMACS with a grating of 150 lines/mm, giving a spectral resolution of $\sim600$ at 7200 $\rm\AA$ for a $0\farcs9$ slit and used OG570 filter to avoid the overlap.
This set-up give the wavelength coverage of 5700 to 9740 $\rm\AA$.
Note that we used chips 5 and 8 of the f/4 camera which have the highest sensitivities among the IMACS CCD chips.
To maximize the S/N, each spectrum was binned by $2\times2$ during the observation.

For data reduction, we followed general reduction processes; bias-subtraction and flat-fielding.
After the wavelength calibration with HeNeAr lines, we generated 2-dimensional maps of sky lines, by performing a polynomial fitting for pixel values along the spatial direction.
We combined the processed 2-dimensional spectra from different chips with the astronomical software SWarp \citep{Bertin10}.
Note that there are CCD gaps along the spectral direction, which are located at $\lambda_{\rm obs}=$ 6530-6630 $\rm \AA$ and $>9700~\rm\AA$.
Identical to the Gemini spectra, the fluxes within a $1\farcs0$ diameter aperture were extracted, and flux-calibrated using both the spectra of A0V standard stars (HD18225, HD85589) and the $i$-magnitude of each target.
The binning was also performed for these spectra in a similar way to the Gemini spectra, but the binned spectra have a spectral resolution of $\sim600$.

\begin{figure*}
\centering
\epsscale{0.9}
\plotone{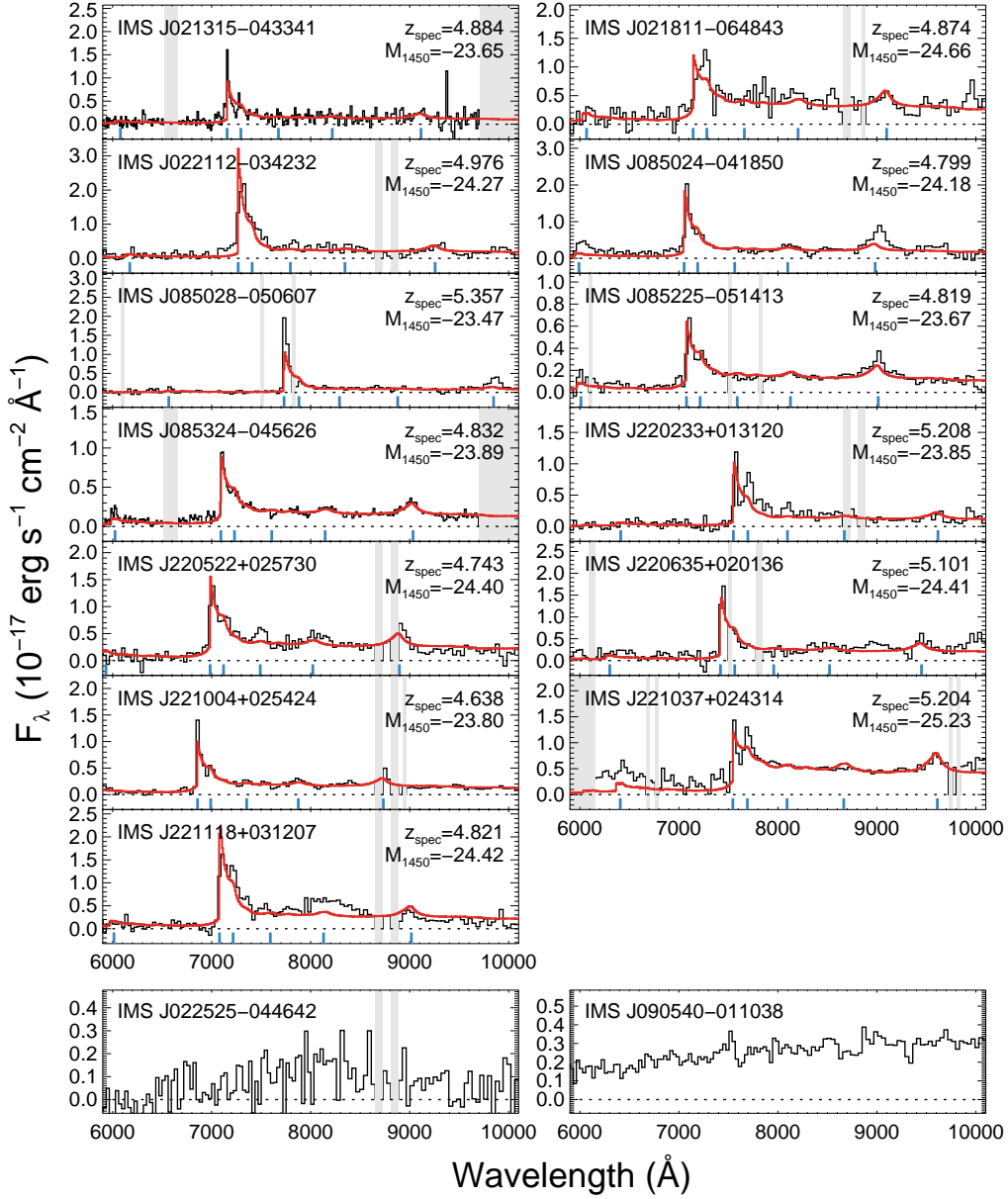}
\caption{
The optical spectra of the identified candidates; the top 13 spectra show the $z\sim5$ quasars, while the bottom 2 are identified as non-quasar objects. The binned spectra are shown as the black solid lines, while the red solid lines are the best-fit models for each quasar. The blue marks indicate the wavelengths of possible emission lines of each quasar (Ly$\beta$, Ly$\alpha$, \ion{N}{5}, \ion{O}{1}, \ion{Si}{4}, and \ion{C}{4}, from short to long wavelengths).
The dotted lines indicate $F_{\lambda}=0$ and the shaded regions represent the bad columns (e.g., hot pixels or gap) on CCD or the wavelength range not covered by the observational configuration.
\label{fig:spec}}
\end{figure*}

\subsection{Supplemental Spectroscopic Redshift Sample} \label{sec:spec_sup}

For some of the medium-band observed objects, we adopted their spectral parameters such as $\zspec$ and $M_{1450}$ from literature.
They mainly come from the catalog of $z\sim5$ quasar candidates by \citetalias{McGreer18} which also used the optical data from CFHTLS to select quasar candidates.
Of the 38 quasars they identified with spectroscopy, we used spectral parameters of 18 quasars; they are located in our survey area (IMS) and satisfy our broad-band color criteria with the magnitude limit ($i\lesssim23$ mag).
Two quasars among them, IMS J221520$-$000908 and IMS J222216$-$000406, are also identified by \cite{Ikeda17},
but we took their spectral parameters from \citetalias{McGreer18}.
Note that we revise the $\zspec$ of IMS J140150$+$514310 from 4.20 in \citetalias{McGreer18} to 5.17 since the Ly$\alpha$ and Ly$\beta$ lines are located at 7500 and 6320$~\rm\AA$, respectively, along with other possible emission lines at the same redshift (see Figure 9 of \citetalias{McGreer18}).
The $M_{1450}$ value of the quasar is also revised with the $\zspec$.
Additionally, we used the spectral parameters of 4 quasars, which are not included in the final catalog of \citetalias{McGreer18} but spectroscopically identified by them.
Consequently, we used the $\zspec$ and $M_{1450}$ values of 22 quasars from \citetalias{McGreer18}, which are listed in Table \ref{tbl:fit}.
Note that there are no $M_{1450}$ values for the 4 quasars excluded in the final catalog of \citetalias{McGreer18}.
Including our spectroscopically identified quasars, the total number of spectroscopically identified quasars we used for our study is 35.

\section{RESULTS}\label{sec:results}

\subsection{Spectroscopic Identification of Quasars} \label{sec:identify}

We present the optical spectra of the 15 broad-band-selected quasar candidates in Figure \ref{fig:spec}. 
13 of them have clear Ly$\alpha$ breaks at 7000-7500 $\rm\AA$ in their spectra,
showing that they are high-redshift quasars.
Most of the quasars also have strong Ly$\alpha$ emission line (S/N $\geq5$), while IMS J021811$-$064843 does not.
In addition, some spectra show broad emission lines such as \ion{C}{4} (e.g., IMS J085024$-$041850, IMS J085324$-$045626, IMS J221037$+$024314).
The quasar spectra we obtained show no significantly unusual feature, except for IMS J221118$+$031207 which has a seemingly broadened Fe complex at $\sim8000~\rm\AA$.
Out of the 15 candidates we observed, ten quasars (marked with ${\ddagger}$ in Table \ref{tbl:broad}) are newly discovered ones,
and three were independently identified by \citetalias{McGreer18}.
On the other hand, the other 2 candidates selected by broad-band color criteria are identified as non-quasar objects
(bottom panels in Figure \ref{fig:spec}), considering that they have no significant break or emission line feature.

\subsection{Medium-band Color Selection and Its Efficiency} \label{sec:coleff}

\begin{figure}
\centering
\epsscale{1.2}
\plotone{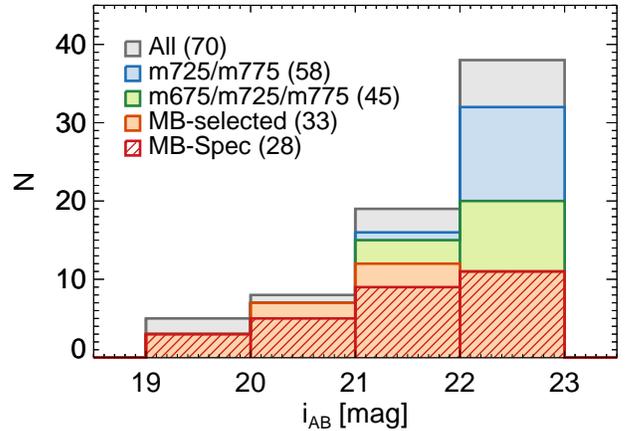}
\caption{
The histogram of our quasar candidates along the $i$-band magnitude.
While the 70 candidates for $z\sim5$ quasar are shown as the gray histogram, 
the total 58 candidates with $m725$- and $m775$-band observations are given in the blue histogram,
and 45 of them also have $m675$-band photometry (the green histogram).
Among the medium-band observed candidates, 33 candidates satisfy the medium-band color criteria given by \cite{Jeon16} (the orange histogram), and 28 of them were spectroscopically identified as high redshift quasars by this work and previous works (\citetalias{McGreer18}; \citealt{Ikeda17}).
Note that the 28 candidates are given in the red hatched histogram for easy distinction from the orange histogram.
There are 7 quasars with spectroscopy, excluded from the red histogram due to the lack of medium-band observations or their medium-band colors unsatisfying the color criteria.
\label{fig:hist}}
\end{figure}

In this section, we examine the effectiveness of using medium-band data obtained by SQUEAN for finding $z\sim5$ quasars.
Figure \ref{fig:hist} summarizes the numbers of our candidates along the $i$-band magnitude at various selection or observation stages.
There are 70 broad-band-selected candidates (gray histogram), 45 of them were observed in three medium-bands ($m675$, $m725$, and $m775$; green histogram), and 33 of the 45 candidates satisfy the color criteria (orange histogram) given by \cite{Jeon16}.
Among the 33 medium-band-selected candidates, 28 of them have spectroscopic data, and all of them are identified as high redshift quasars (red histogram).
We suggest that the other 5 medium-band-selected candidates are also high redshift quasars that they are bright ($i<22$ mag) and have high S/N medium-band data and yet their SED shape is very much in agreement with the other confirmed quasars.
On the other hand, 27 \% of (12 out of 45) of the broad-band-selected candidates were removed by the medium-band color criteria.
Out of the 12 excluded candidates, four turned out to be quasars.
IMS J143945$+$562627 and IMS J221004$+$025424 are excluded due to their redshift ($z\leq4.7$),
so their exclusion is under special circumstances.
The other two, IMS J220522$+$025730 and IMS J220635$+$020136, are not selected since they have shallow depth images in the $m675$-band, which gives only a lower limit on the $m675-m725$ color.
Excluding these two quasars, we estimate that the contaminants occupy 23 \% (10 out of 43) of the broad-band-selected sample.
Note that we assumed that the 10 candidates are all non-quasars or quasars that are out of the explored redshift range. 
Figure \ref{fig:hist} shows the histogram of our candidates for $z\sim5$ quasar along the $i$-band magnitude.
The medium-band selection becomes more important if we concentrate on faint objects.
At $22<i<23$, in comparison to $i<23$, the contamination rate increases to 47 \% (9 out of 19, except IMS J220635$+$020136), for the broad-band selected candidates that are rejected after the medium-band observation.
It is due to the increase of faint red stars that can act as interlopers,
and without the medium-band approach, the exclusion of such objects become more challenging as we go to fainter magnitudes.
Consequently, this medium-band-approach is an effective way to narrow down the number of plausible candidates for $z\sim5$ quasars.

However, our method is limited by the broad-band selection and photometry.
As one can see in Figure \ref{fig:ccd}, there are 4 quasars at $z\sim5$ reported by \citetalias{McGreer18} that were excluded from our broad-band-selected candidates (purple open diamonds).
Except for a quasar with a red $i-z$ color of 1.0, not included in the final catalog of \citetalias{McGreer18}, the other three quasars were not selected by our selection criteria because there are small differences in broad-band magnitudes ($\sim0.1$ mag) between \citetalias{McGreer18} and this work.
In other words, we may have missed 10 \% (4 out of 39) of quasars (or candidates) during our broad-band selection.
We checked if the photometric accuracy is the main reason for missing 10 \% of quasars during the broad-band selection by using our SED model described in \ref{sec:zdet}.
We randomly generated $10^5$ mock quasars at $4.7 \leq z \leq 5.4$ based on the SED model, controlled by the QLF of \citetalias{McGreer18} with the parameter ranges determined by previous studies (see details in Section \ref{sec:zdet}), including photometric uncertainties of 0.1 mag.
11.4 \% of the mock quasars are rejected by our criteria, corresponding to the fraction of the missed quasars.
Thus, to have a highly complete sample, a rather generous broad-band selection or a selection from a sample with higher photometry accuracy is desirable before applying the medium-band selection.

\begin{figure*}
\centering
\epsscale{0.9}
\plotone{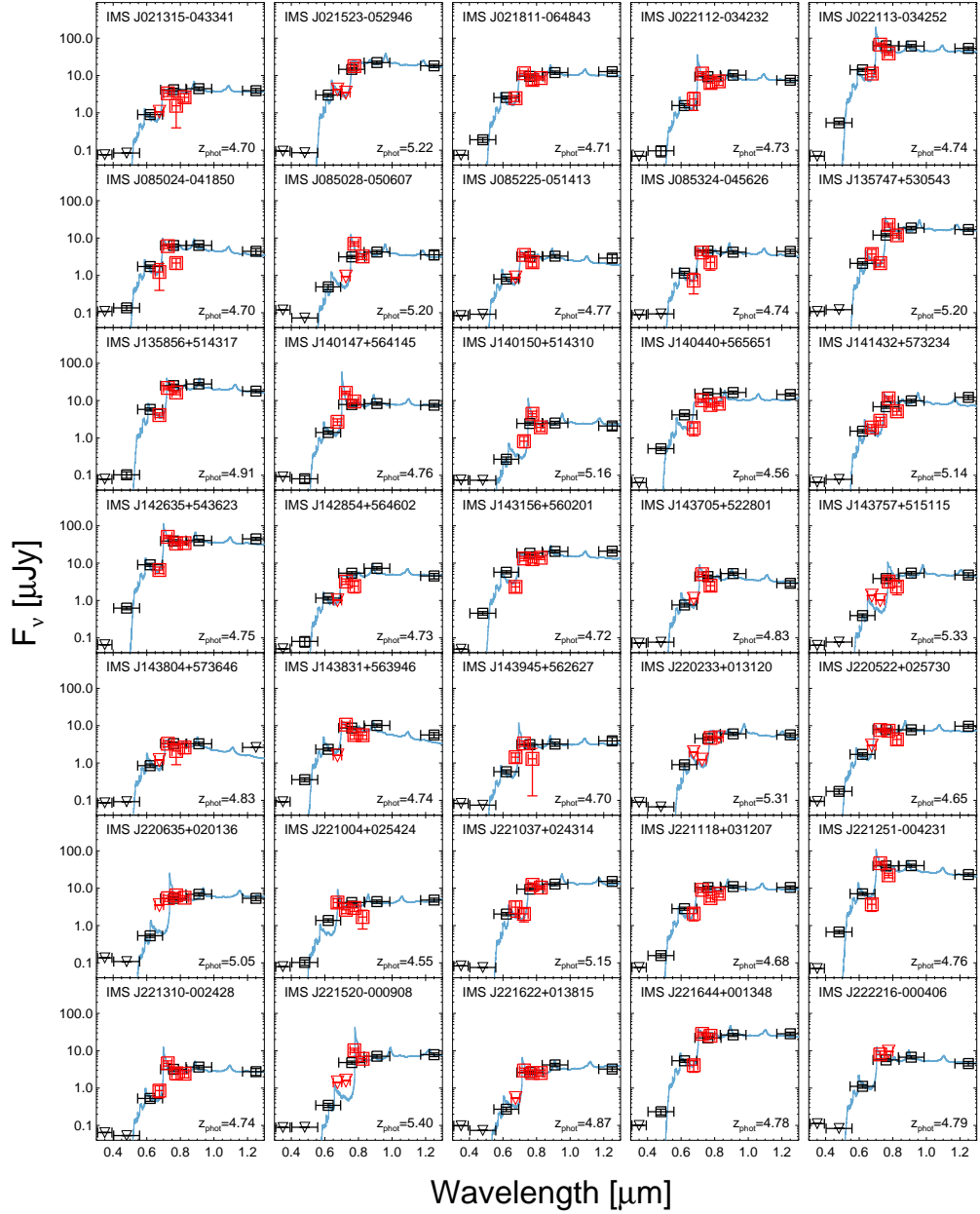}
\caption{
The SEDs of quasars with broad- ($ugrizJ$) and medium-band ($m675$-$m825$) fluxes, 
which are shown as the black and red squares, respectively.
Note that the downside triangles represent the $2.7\sigma$ upper limits.
The best-fit model of each quasar is shown with the blue solid line, 
for which $\zphot$ values are also indicated in each panel.
\label{fig:sed}}
\end{figure*}

\subsection{SED-fitting and Redshift Measurements} \label{sec:zdet}

The estimation of $\zspec$ requires spectra with good S/N, which is usually expensive in observing time. 
As a good alternative, $\zphot$ does not require observing time as extensive as spectroscopy, 
and is still useful for deriving properties of high redshift quasars. 
While $\zphot$ of quasars can be determined by red colors from a sharp break at wavelength shorter than Ly$\alpha$, their accuracy depends critically on how exactly one can sample the break in multi-band photometry. 
In that regard, medium-band photometry can be useful since its dense wavelength sampling can improve the wavelength estimation of the break.
We describe here our derivation of $\zphot$ and $\zspec$ with a quasar SED model.

\subsubsection{Quasar SED Model \label{sec:modelQ}}

We generated an artificial quasar SED model based on the composite spectrum of SDSS quasars \citep{Vanden01}.
Note that there is a more recent composite spectrum of SDSS quasars without the effect of host galaxy contamination \citep{Selsing16}.
But the rest-frame wavelength coverage is only $\lambda_{\rm rest}>1000~\rm\AA$ for that template ($\lambda_{\rm rest}>800~\rm\AA$ for \citealt{Vanden01}) and the host contamination is not a significant factor at rest-frame UV wavelengths for a quasar with $L_{\rm bol}\gtrsim10^{46}$ erg s$^{-1}$  \citep{Shen11}, which is comparable to our quasars.
Based on the spectra, we used spectral parameters described below to generate our quasar SED models for fitting.

The quasar continuum slope of the SDSS composite spectrum is $\alpc=-1.54$ \citep{Vanden01}, 
where $F_{\lambda}\propto \lambda^{\alpc}$.
Note that, in a wavelength range of 1450 to 2200 $\rm\AA$, $\alpc$ ranges from $-2.5$ to $-0.5$ \citep{Davis07,Shen11,Mazzucchelli17}.
To change the continuum slope of the composite spectrum for a given $\alpc$, we multiplied a factor of $(\lambda/1000~\rm{\AA})^{\alpc-\alpha_{\lambda,0}}$ to the composite spectrum, where $\alpha_{\lambda,0}=-1.54$.

The equivalent width of Ly$\alpha$ and \ion{N}{5} $\lambda1240$ (hereafter EW) is also important to determine the shape of quasar SED model.
For the EW estimation, we integrated the Ly$\alpha$ and \ion{N}{5} fluxes over the continuum fluxes 
at the range of $1160 \leq \lambda_{\rm rest}~({\rm \AA}) \leq 1290$ ($f_{\rm Ly\alpha+NV}$).
In order to adjust the EW value of the composite spectrum to an arbitrary EW value, 
we scaled the $f_{\rm Ly\alpha+NV}$ at that wavelength range by adjusting the power of $p$;
$f_{\rm Ly\alpha+NV}=f_{\rm Ly\alpha+NV,V01} \times ( \lambda_{\rm rest}/1290~{\rm\AA} )^{p}$,
where $f_{\rm Ly\alpha+NV,V01}$ is the flux measured from the original spectrum of \cite{Vanden01}.

After adjusting the $\alpc$ and EW, we applied IGM attenuation to the composite spectra,
using the polynomial approximation in \cite{Madau96}.
The effective optical depth for the Ly$\alpha$ emission line at $4.5<z<5.5$ is in line with the values based on several observations \citep{Songaila04,Fan06} and other simulated templates for $z\sim5$ quasars (\citealt{McGreer13}; \citetalias{McGreer18}).

Including $M_{1450}$ as a scaling factor, in summary, four parameters ($z$, $M_{1450}$, $\alpc$, and EW) are used to generate our quasar models for the fitting.
Note that the $M_{1450}$ and EW are left as independent parameters for the fitting instead of adopting the Baldwin effect, the correlation between EWs of quasar emission lines and the continuum luminosities \citep{Baldwin77}, considering the uncertainty of the Baldwin effect for Ly$\alpha$ at high redshift \citep{Constantin02,Dietrich02}.
Several quasar model tracks from $z=4.5$ to 5.5 are shown as the gray dots with solid lines in Figures \ref{fig:ccd} and \ref{fig:ccd2}, where we adopted $M_{1450}=-24$ mag, $\alpc=-1.6$ and $\log {\rm (EW/\AA)}=1.5$.
Our simulated models also satisfy the criteria, given by \cite{McGreer13} and \cite{Jeon16}.

\subsubsection{Photometric Redshift\label{sec:zphot}}

Based on the fluxes from the broad- and the medium-band observations, $\zphot$ was determined by finding minimum $\chi^{2}$ value between the observed fluxes and the model fluxes, where $\chi^{2}$ is defined as

\begin{equation}
\chi^{2}=\sum_{i} \chi^{2}_{i}+ \sum_{j}\chi^{2}_{j} \label{eq:chi2}.
\end{equation}

\noindent $\chi^{2}_{i}$, the first term is a standard form of $\chi^{2}$ for the filters with detection,

\begin{equation}
\chi^{2}_{i}=\left(\frac{f_{o,i}-f_{m,i}}{\sigma_{i}}\right)^{2} \label{eq:chi2i},
\end{equation}

\noindent where $f_{o,i}$ is the observed flux in the $i$-th band, $\sigma_{i}$ is the standard deviation (or uncertainty) of the observed flux, and $f_{m,i}$ is the model flux in the same band, which is calculated by integrating the quasar model fluxes with the weight of the transmission curve of the band.
For the case of the filters with upper limit of fluxes, we refer to the $\chi^{2}$ derivation by \cite{Sawicki12},
which gives $\chi^{2}_{j}$ of the second term of Eq. (\ref{eq:chi2}),

\begin{equation}
\begin{aligned}
\chi^{2}_{j} &=-2 \ln \int_{-\infty}^{f_{lim,j}} \exp \left[-\frac{1}{2}\left(\frac{f_{o,j}-f_{m,j}}{\sigma_{j}}\right)^{2}\right] df \\
&=-2 \ln \Bigg\{ \sqrt{\frac{\pi}{2}}~\sigma_{j} \Bigg[ 1 + {\rm erf} \left(\frac{f_{lim,j}-f_{m,j}}{\sqrt{2}\sigma_{j}}\right) \Bigg] \Bigg\} \label{eq:chi2j}
\end{aligned}
\end{equation}

\begin{deluxetable*}{lcccccccc}
\tablewidth{700pt}
\tabletypesize{\scriptsize}
\tablecaption{Quantities of $z\sim5$ Quasars from the Model Fitting \label{tbl:fit}}
\tablehead{
\colhead{} & \multicolumn{4}{c}{Photometry} & \multicolumn{4}{c}{Spectroscopy } \\
\cmidrule(lr){2-5} \cmidrule(lr){6-9}
\colhead{ID\hspace{5cm}}  & \colhead{$\zphot$} & \colhead{$M_{1450}$} & \colhead{$\alpc$} & \colhead{log EW} & \colhead{$\zspec$} & \colhead{$M_{1450}$} & \colhead{log EW} & \colhead{\hspace{0.3cm}Ref.}\hspace{0.3cm} \\
\colhead{}  & \colhead{} & (mag) & \colhead{} & ($\rm\AA$) &\colhead{} & (mag) & ($\rm\AA$) &
}
\startdata
IMS J021315$-$043341\hspace{2cm}  & $4.70^{+0.23}_{-0.10}$ & $-23.7^{+  0.1}_{-  0.2}$ & $-1.8^{+ 0.5}_{- 0.9}$ & $0.9^{+0.8}_{-0.3}$ & $4.884^{+0.003}_{-0.035}$ & $-23.65^{+  0.73}_{-  0.45}$ & $1.9^{+0.3}_{-0.3}$ & (1)  \\
IMS J021523$-$052946  & $5.22^{+0.17}_{-0.07}$ & $-25.8^{+  0.1}_{-  0.1}$ & $-2.6^{+ 0.4}_{- 0.6}$ & $\lesssim 0.5$ & 5.13 & $-25.6$ & -  & (2)  \\
IMS J021811$-$064843  & $4.71^{+0.04}_{-0.07}$ & $-24.8^{+  0.1}_{-  0.1}$ & $-1.8^{+ 0.5}_{- 0.5}$ & $0.9^{+0.7}_{-0.3}$ & $4.874^{+0.033}_{-0.028}$ & $-24.66^{+  0.23}_{-  0.20}$ & $1.6^{+0.2}_{-0.6}$ & (1)  \\
IMS J022112$-$034232  & $4.73^{+0.17}_{-0.03}$ & $-24.5^{+  0.1}_{-  0.2}$ & $-1.8^{+ 0.4}_{- 0.7}$ & $1.9^{+0.2}_{-0.6}$ & $4.976^{+0.003}_{-0.003}$ & $-24.27^{+  0.23}_{-  0.14}$ & $2.2^{+0.2}_{-0.2}$ & (1)  \\
IMS J022113$-$034252  & $4.74^{+0.03}_{-0.01}$ & $-26.4^{+  0.1}_{-  0.1}$ & $-1.8^{+ 0.4}_{- 0.2}$ & $1.9^{+0.2}_{-0.2}$ & 5.02 & $-27.0$ & -  & (2)  \\
IMS J085024$-$041850  & $4.70^{+0.07}_{-0.14}$ & $-24.1^{+  0.1}_{-  0.1}$ & $-2.8^{+ 1.0}_{- 0.7}$ & $1.3^{+0.4}_{-0.7}$ & $4.799^{+0.003}_{-0.003}$ & $-24.18^{+  0.07}_{-  0.08}$ & $2.0^{+0.2}_{-0.2}$ & (1)  \\
IMS J085028$-$050607  & $5.20^{+0.17}_{-0.04}$ & $-23.9^{+  0.2}_{-  0.1}$ & $-2.4^{+ 1.0}_{- 1.1}$ & $1.7^{+0.2}_{-0.2}$ & $5.357^{+0.003}_{-0.008}$ & $-23.47^{+  0.22}_{-  0.11}$ & $2.2^{+0.2}_{-0.2}$ & (1)  \\
IMS J085225$-$051413  & $4.77^{+0.20}_{-0.09}$ & $-23.5^{+  0.2}_{-  0.1}$ & $-2.8^{+ 1.0}_{- 0.7}$ & $1.1^{+0.6}_{-0.5}$ & $4.819^{+0.003}_{-0.003}$ & $-23.67^{+  0.08}_{-  0.08}$ & $1.7^{+0.2}_{-0.2}$ & (1)  \\
IMS J085324$-$045626  & $4.74^{+0.21}_{-0.10}$ & $-23.8^{+  0.1}_{-  0.2}$ & $-2.2^{+ 0.7}_{- 0.8}$ & $1.1^{+0.6}_{-0.5}$ & $4.832^{+0.004}_{-0.004}$ & $-23.89^{+  0.04}_{-  0.05}$ & $1.8^{+0.2}_{-0.2}$ & (1)  \\
IMS J135747$+$530543  & $5.20^{+0.10}_{-0.03}$ & $-25.5^{+  0.1}_{-  0.1}$ & $-2.0^{+ 0.4}_{- 0.5}$ & $1.5^{+0.2}_{-0.2}$ & 5.32 & $-25.5$ & -  & (2)  \\
IMS J135856$+$514317  & $4.91^{+0.04}_{-0.04}$ & $-25.7^{+  0.1}_{-  0.1}$ & $-2.4^{+ 0.4}_{- 0.3}$ & $1.3^{+0.2}_{-0.2}$ & 4.97 & $-25.9$ & -  & (2)  \\
IMS J140147$+$564145  & $4.76^{+0.06}_{-0.02}$ & $-24.5^{+  0.1}_{-  0.1}$ & $-1.8^{+ 0.2}_{- 0.5}$ & $2.1^{+0.2}_{-0.2}$ & 4.98 & $-24.7$ & -  & (2)  \\
IMS J140150$+$514310  & $5.16^{+0.15}_{-0.01}$ & $-23.4^{+  0.2}_{-  0.1}$ & $-2.0^{+ 0.6}_{- 0.9}$ & $1.9^{+0.2}_{-0.2}$ & 5.17$^{a}$ & $-23.4^{a}$ & -  & (2)  \\
IMS J140440$+$565651  & $4.56^{+0.09}_{-0.03}$ & $-24.7^{+  0.1}_{-  0.1}$ & $-1.6^{+ 0.4}_{- 0.8}$ & $\lesssim 0.5$ & 4.74 & -  & -  & (2)  \\
IMS J141432$+$573234  & $5.14^{+0.04}_{-0.07}$ & $-24.8^{+  0.1}_{-  0.1}$ & $-2.4^{+ 0.9}_{- 0.8}$ & $1.3^{+0.3}_{-0.7}$ & 5.16 & $-24.7$ & -  & (2)  \\
IMS J142635$+$543623  & $4.75^{+0.01}_{-0.01}$ & $-26.2^{+  0.1}_{-  0.1}$ & $-2.0^{+ 0.2}_{- 0.5}$ & $1.7^{+0.2}_{-0.2}$ & 4.76 & $-26.3$ & -  & (2)  \\
IMS J142854$+$564602  & $4.73^{+0.27}_{-0.12}$ & $-24.0^{+  0.1}_{-  0.3}$ & $-1.8^{+ 0.5}_{- 1.1}$ & $\lesssim 0.5$ & 4.73 & $-24.0$ & -  & (2)  \\
IMS J143156$+$560201  & $4.72^{+0.04}_{-0.04}$ & $-25.3^{+  0.1}_{-  0.1}$ & $-2.2^{+ 0.7}_{- 0.6}$ & $\lesssim 0.5$ & 4.75 & -  & -  & (2)  \\
IMS J143705$+$522801  & $4.83^{+0.14}_{-0.11}$ & $-23.8^{+  0.2}_{-  0.1}$ & $-2.4^{+ 0.7}_{- 0.5}$ & $1.5^{+0.4}_{-0.5}$ & 4.78 & -  & -  & (2)  \\
IMS J143757$+$515115  & $5.33^{+0.12}_{-0.24}$ & $-24.2^{+  0.3}_{-  0.1}$ & $-2.0^{+ 1.0}_{- 0.7}$ & $1.5^{+0.3}_{-0.9}$ & 5.17 & $-24.1$ & -  & (2)  \\
IMS J143804$+$573646  & $4.83^{+0.20}_{-0.20}$ & $-23.5^{+  0.2}_{-  0.2}$ & $\lesssim -3.6$ & $0.7^{+1.0}_{-0.2}$ & 4.84 & $-23.5$ & -  & (2)  \\
IMS J143831$+$563946  & $4.74^{+0.04}_{-0.07}$ & $-24.5^{+  0.1}_{-  0.1}$ & $\lesssim -3.6$ & $1.1^{+0.5}_{-0.5}$ & 4.82 & -  & -  & (2)  \\
IMS J143945$+$562627  & $4.70^{+0.03}_{-0.06}$ & $-23.4^{+  0.1}_{-  0.1}$ & $-1.4^{+ 0.8}_{- 0.8}$ & $1.9^{+0.2}_{-0.9}$ & 4.70 & $-23.2$ & -  & (2)  \\
IMS J220233$+$013120  & $5.31^{+0.10}_{-0.23}$ & $-24.5^{+  0.2}_{-  0.1}$ & $-2.8^{+ 1.0}_{- 0.7}$ & $\lesssim 0.5$ & $5.208^{+0.022}_{-0.003}$ & $-23.85^{+  0.10}_{-  0.13}$ & $2.0^{+0.2}_{-0.2}$ & (1)  \\
IMS J220522$+$025730  & $4.65^{+0.07}_{-0.07}$ & $-24.4^{+  0.1}_{-  0.1}$ & $-1.8^{+ 0.5}_{- 0.6}$ & $1.3^{+0.4}_{-0.7}$ & $4.743^{+0.004}_{-0.012}$ & $-24.40^{+  0.15}_{-  0.12}$ & $1.8^{+0.2}_{-0.2}$ & (1)  \\
IMS J220635$+$020136  & $5.05^{+0.07}_{-0.15}$ & $-24.2^{+  0.1}_{-  0.1}$ & $-1.4^{+ 0.3}_{- 0.5}$ & $1.9^{+0.2}_{-0.3}$ & $5.101^{+0.003}_{-0.003}$ & $-24.41^{+  0.11}_{-  0.08}$ & $1.9^{+0.2}_{-0.2}$ & (1)  \\
IMS J221004$+$025424  & $4.55^{+0.07}_{-0.05}$ & $-23.6^{+  0.1}_{-  0.1}$ & $-1.2^{+ 0.4}_{- 0.7}$ & $1.7^{+0.4}_{-0.7}$ & $4.638^{+0.003}_{-0.004}$ & $-23.80^{+  0.06}_{-  0.05}$ & $1.8^{+0.2}_{-0.2}$ & (1)  \\
IMS J221037$+$024314  & $5.15^{+0.07}_{-0.06}$ & $-25.2^{+  0.1}_{-  0.1}$ & $-1.8^{+ 0.6}_{- 0.7}$ & $0.9^{+0.6}_{-0.3}$ & $5.204^{+0.010}_{-0.012}$ & $-25.23^{+  0.03}_{-  0.03}$ & $1.4^{+0.2}_{-0.2}$ & (1)  \\
IMS J221118$+$031207  & $4.68^{+0.06}_{-0.12}$ & $-24.7^{+  0.2}_{-  0.1}$ & $-1.8^{+ 0.6}_{- 0.5}$ & $0.7^{+0.8}_{-0.2}$ & $4.821^{+0.003}_{-0.003}$ & $-24.42^{+  0.12}_{-  0.13}$ & $2.0^{+0.2}_{-0.2}$ & (1)  \\
IMS J221251$-$004231  & $4.76^{+0.03}_{-0.01}$ & $-26.0^{+  0.1}_{-  0.1}$ & $-2.6^{+ 0.2}_{- 0.3}$ & $1.7^{+0.2}_{-0.2}$ & 4.95 & $-26.3$ & -  & (3)  \\
IMS J221310$-$002428  & $4.74^{+0.20}_{-0.03}$ & $-23.4^{+  0.1}_{-  0.2}$ & $-1.8^{+ 0.8}_{- 0.8}$ & $1.9^{+0.3}_{-0.5}$ & 4.80 & $-23.5$ & -  & (2)  \\
IMS J221520$-$000908  & $5.40^{+0.06}_{-0.20}$ & $-24.5^{+  0.1}_{-  0.1}$ & $-1.2^{+ 0.5}_{- 0.5}$ & $2.1^{+0.2}_{-0.2}$ & 5.28 & $-24.5$ & -  & (2)  \\
IMS J221622$+$013815  & $4.87^{+0.20}_{-0.12}$ & $-23.4^{+  0.1}_{-  0.2}$ & $-0.8^{+ 0.3}_{- 0.7}$ & $1.7^{+0.2}_{-0.4}$ & 4.93 & $-23.3$ & -  & (2)  \\
IMS J221644$+$001348  & $4.78^{+0.09}_{-0.04}$ & $-25.8^{+  0.1}_{-  0.1}$ & $-1.8^{+ 0.4}_{- 0.4}$ & $1.3^{+0.3}_{-0.7}$ & 5.01 & $-25.8$ & -  & (2)  \\
IMS J222216$-$000406  & $4.79^{+0.24}_{-0.10}$ & $-24.2^{+  0.2}_{-  0.2}$ & $-2.2^{+ 0.5}_{- 0.6}$ & $1.5^{+0.6}_{-0.9}$ & 4.95 & $-24.3$ & -  & (2)  \\
\enddata
\tablecomments{
The systematic uncertainty of the redshift determination with the Ly$\alpha$ fitting
($\Delta z \lesssim 0.1$; \citealt{Kim15,Kim18}; \citetalias{McGreer18}) is not included in the uncertainties of $\zphot$ and $\zspec$.
The spectral properties are from (1) this work, (2) \citetalias{McGreer18}, and (3) \cite{McGreer13}.
For spectroscopic data in this work, we fixed $\alpc$ to $-1.54$ when fitting our quasar SED model (see Section \ref{sec:zspec}).
Note that $M_{1450}$ from (2) and (3) are determined by the $i$-band magnitudes and $\zspec$, which are matched to model quasar spectra. The difference in cosmological parameters between the literature and this work is also concerned.
}
\tablenotetext{a}{For IMS J140150+514310, \citetalias{McGreer18} provides $\zspec=4.20$. 
But, we revise it to be $\zspec=5.17$ from the Ly$\alpha$ break in the spectrum shown in Fig. 9 of \citetalias{McGreer18} (see details in Section \ref{sec:spec_sup}).
$M_{1450}$ is the value that assumes $\zspec=5.17$.}
\end{deluxetable*}

\noindent where $f_{lim,j}$ is the upper limit of the flux in the $j$-th band, $f_{m,j}$ is the model flux in the same band, $\sigma_{j}$ is the sensitivity in the same band, and ${\rm erf}(x)$ is the error function for the numerical calculation; ${\rm erf}(x)=(2/\sqrt{\pi})\int_{0}^{x}e^{-t^2}dt$.
Note that we limited the $\chi^{2}_{j}$ value by $\chi^{2}_{j} \leq 0$ to restrict the $\chi^{2}$ value being negative.

\begin{figure*}[ht]
\centering
\epsscale{1.0}
\plotone{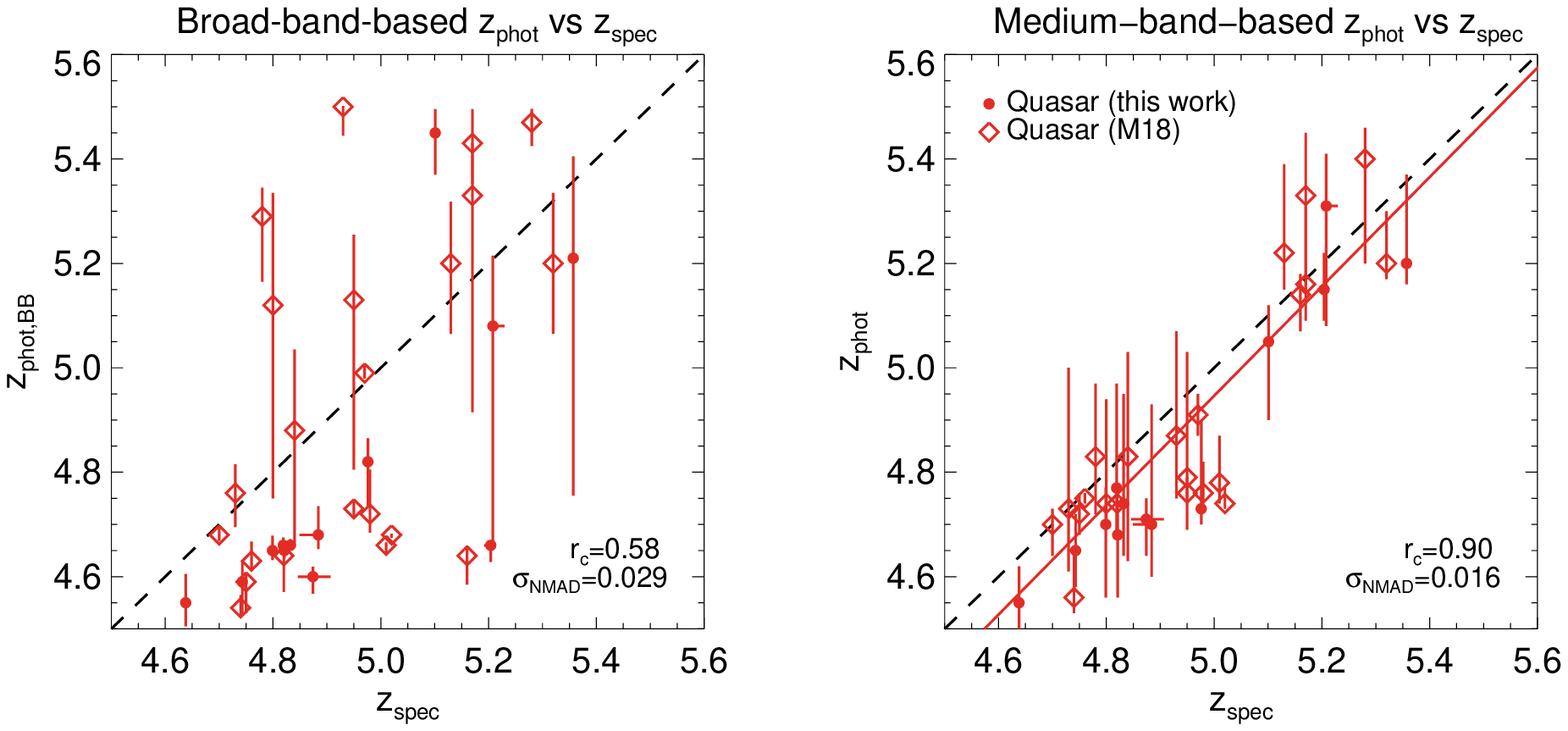}
\caption{
Comparison of $\zphot$ versus $\zspec$ of quasars at $z\sim5$ for the $\zphot$ values derived from broad-band photometry only (left) and the $\zphot$ values from the broad- and medium-band data (right).
The symbols are the same as those in Figure \ref{fig:field}.
The black dotted line shows the case where $\zphot$ is identical to $\zspec$, 
and the red solid line indicates the best-fit result.
The Pearson correlation coefficient ($r_{c}$) and the scatter of normalized median absolute deviation ($\sigma_{\rm NMAD}$) are remarked.
\label{fig:compz}}
\end{figure*}

The minimum $\chi^{2}$ was searched in the following parameter space of $z$, $M_{1450}$, $\alpc$, and EW;
$4.5\leq z \leq5.5$ with a step size of 0.01, $-27.5\leq M_{1450}~({\rm AB~mag})\leq-22.5$ with a step size of 0.1 mag, $-3.6\leq \alpc \leq 0.4$ with a step size of 0.2, and 0.5 $\leq \log($EW$/{\rm \AA}) \leq 2.5$ with a step size of 0.2.
Note that the above ranges of $\alpc$ and EW are chosen to cover the $\alpc$ and EW values within about 2-$\sigma$ of the average values for high redshift quasars that have the values of $\alpc=-1.6\pm1.0$ \citep{Mazzucchelli17} and $\log($EW$/\rm\AA)$ of $=1.542 \pm 0.391$ in the rest-frame \citep{Banados16}. For each model, we estimated the model flux in each band by calculating the mean flux in each band, which was weighted by the filter transmission curve.

For each quasar, we calculated $\chi^{2}_{\rm red}$ value (the reduced $\chi^{2}$ value, defined as $\chi^{2}_{\rm red} \equiv \chi^{2}/\nu_{\rm dof}$, where the $\nu_{\rm dof}$ is the degree of freedom) for each model with broad ($grizJ$) and existing medium-band ($m675$-$m825$) fluxes.
For the broad-band photometry, we gave additional errors on the broad-band magnitudes considering the possible variability of quasars between the observing dates of the broad- and medium-band observations\footnote{
While the CFHTLS and the IMS data were obtained in 2003-2008 and 2009-2013, respectively, 
the medium-band observations were carried out in 2015-2018, 
corresponding to a term of 1-2 yr between the observations in the rest-frame.
The rest-frame far-UV variability of low redshift quasars over a year scale is $\sim0.5$ mag yr$^{-1}$ for the most significant variable fraction of $\sim10~\%$ \citep{Welsh11}.
Therefore, we gave an arbitrary error of 0.1 mag (1-2 yr $\times$ 0.5 mag yr$^{-1}$ $\times$ 10 \% $\sim0.1$ mag) to each broad-band magnitude.
}.
We found the minimum $\chi^{2}_{\rm red}$ value ($\chi^{2}_{\rm red,min}$) as the best-fit result,
and interpolated $\chi^{2}_{\rm red}$ values in the four parameter spaces to find points of $\chi^{2}_{\rm red} = \chi^{2}_{\rm red,min}+1$,
which are regarded as the marginal points for the errors of each parameter at $1\sigma$ confidence level.
Note that the interpolation may over/underestimate the $1\sigma$ errors by the bin size, but we expect that the effect is negligible.
The best-fit results for 35 spectroscopically identified quasars are listed in Table \ref{tbl:fit}, and Figure \ref{fig:sed} shows the SEDs of the quasars with the best-fit models (blue solid lines).

\subsubsection{Spectroscopic Redshift ($\zspec$) \label{sec:zspec}}

Similarly to the broad-band and medium-band SED fit, $\zspec$ and the SED parameters of 13 quasars were also obtained by finding the minimum $\chi^{2}_{\rm red}$ with Eq. (\ref{eq:chi2}) \& (\ref{eq:chi2i}),
but Eq. (\ref{eq:chi2j}) for the upper limit case is not used.
The wavelength range of the fitting was limited to $1100 \leq  \lambda_{\rm obs}~({\rm \AA}) /(1+z_{\rm vis}) \leq 1600$,
where $z_{\rm vis}$ is the redshift determined by visual inspection of the Ly$\alpha$ line on the spectra.
It covers the Ly$\alpha$ line and the quasar continuum for the fitting.
Among the SED parameters, $\alpc$ was fixed to $-1.54$ since the wavelength coverage of our spectra is too narrow to reliably estimate the quasar continuum slope.
In addition, the adopted parameter grid resolution is higher than the case of $\zphot$ when estimating the best-fit parameters and their errors;
the step sizes of $\zspec$, $M_{1450}$, and $\log \rm{EW}$ were pushed down to 0.001, 0.01, and 0.1, respectively.
Note that the systematic uncertainty in $\zspec$ due to the adopted finite grid size is only $\sim0.002$-0.004 for our binned spectra.

In Figure \ref{fig:spec}, the best-fit models are over-plotted with the red lines, and we marked the wavelengths of possible emission lines, such as Ly-$\beta$ $\lambda1025$, Ly$\alpha$, \ion{N}{5} $\lambda1240$, \ion{O}{1} $\lambda1304$, \ion{Si}{4} $\lambda1396$, and \ion{C}{4} $\lambda1549$, with the blue vertical lines.
Also, the best-fit results are listed in Table \ref{tbl:fit}.

\subsubsection{Medium-band Photometric Redshift Accuracy} \label{sec:accuracy}

\begin{figure}
\centering
\epsscale{1.2}
\plotone{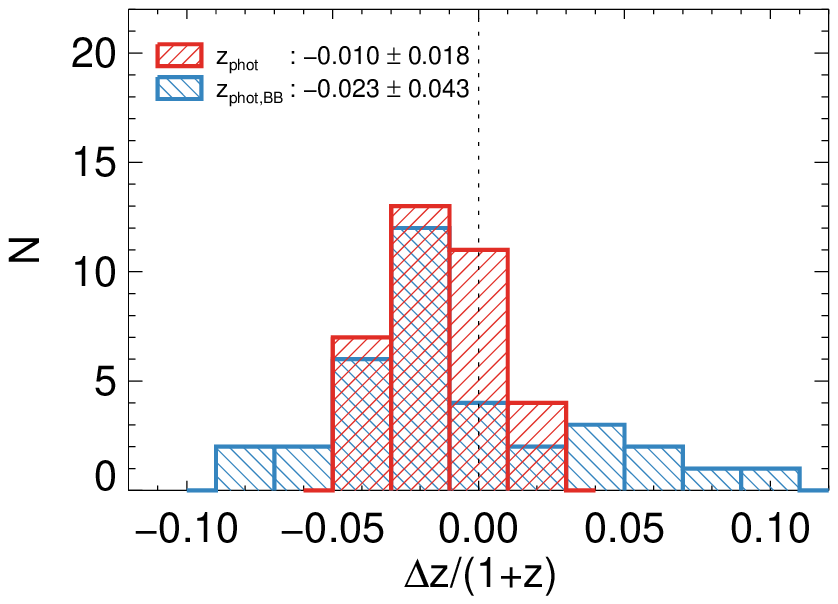}
\caption{
The histogram of $\Delta z / (1+z)$ of $z\sim5$ quasars, where $\Delta z = \zphot - \zspec$.
The red histogram represents the $\Delta z / (1+z)$ distribution based on $\zphot$ including the medium-band photometry,
while the blue one shows that of $z_{\rm phot,BB}$ with only the broad-band photometry.
Their median and standard-deviation values are given in the legend.
The vertical dotted line indicates $\Delta z / (1+z)=0$.
\label{fig:compzh}}
\end{figure}

In Table \ref{tbl:fit}, the best-fit results of our $z\sim5$ quasar sample are listed.
The median uncertainty of the $\zspec$ is only 0.004,
while that of the $\zphot$ is 0.09.
The left panel of Figure \ref{fig:compz} shows the comparison of $\zspec$ and $z_{\rm phot,BB}$,
the photometric redshift determined with only broad-band photometry, for 35 quasars.
They show a loose correlation with a linear Pearson correlation coefficient of $r_{c}=0.58$.
If we introduce the additional medium-band photometry for the $\zphot$ determination,
there is a tight correlation between $\zspec$ and $\zphot$ with the improved $r_{c}$ of 0.90
(the right panel of Figure \ref{fig:compz}).
For the two cases, the scatter of normalized median absolute deviations of $|\Delta z |/ (1+z)$ ($\sigma_{\rm NMAD}$) are 0.029 and 0.016, respectively, where $\Delta z \equiv \zspec - \zphot$ and the $\zspec$ are used for the reference redshifts.

Compared to the identical line (the black dashed line), 
there is a trend of $\zphot$ slightly lower than $\zspec$, which is described by the linear relation of $\zspec = 1.087\times \zphot-0.506$ (the red solid line in Figure \ref{fig:compz}).
For a simple comparison, we plotted the distribution of $\Delta z / (1+z)$ in Figure \ref{fig:compzh}.
The median $\Delta z / (1+z)$ values for $\zphot$ (the red histogram) and $z_{\rm phot,BB}$ (the blue histogram) are slightly biased toward lower redshift ($-0.010$ and $-0.023$, respectively).
The small systematic bias in $\Delta z / (1+z)$ could be explained by the limitation in our quasar models and the filter system.
A quasar model with a stronger Ly$\alpha$ emission can give a $\zphot$ value that is slightly larger than a model with a weaker Ly$\alpha$ emission since both the models give the same amount of flux within a certain passband that samples the light above the sharp break at Ly$\alpha$. 
For that reason, the $\zphot$ probability distribution has a longer tail toward higher redshift.
Since we adopt $\zphot$ at the maximum probability (the best-fit value), this can result in a slight underestimation in $\zphot$.
In addition, the magnitudes at wavelengths longer than Ly$\alpha$ have smaller uncertainties than the wavelength below Ly$\alpha$, and this can lead to a slight underestimation in in $\zphot$ by giving more weight to the longer wavelength magnitudes during the model fitting.
Then the fitting procedure tries to fit the longer wavelength magnitudes better by adjusting the Ly$\alpha$ strength to preferentially allow strong Ly$\alpha$ emission model with larger $\zphot$ values.
We confirm this by increasing the photometry accuracy of a filter below Ly$\alpha$ of a $\zphot=4.7$ quasar from 0.05 to 0.5 mag.
When the photometric error increases, the $\zphot$ value drops by 0.1.
Previous studies of quasar observations in medium-bands also support this explanation.
\cite{Jeon16} used similar models that have a sharp break to measure $\zphot$ of the bright quasar sample at $4.7<z<6.0$
with the SQUEAN medium-band observations, and the $\Delta z / (1+z)$ distribution is a Gaussian distribution of $-0.010\pm0.012$ \citep{Jeon16}.
On the other hand, \cite{Wolf03} used the SDSS quasar spectrum \citep{Vanden01} without IGM attenuation for the $\zphot$ determination of low redshift quasars at $0.6<z<3.5$ with medium-band observations from the COMBO-17 survey.
Their $\zphot$ values are almost identical to $\zspec$ with uncertainty of $\lesssim 0.05$, corresponding to the low IGM attenuation toward the lower redshift quasars.

The standard deviation of the $\zphot$ case (0.018) is smaller than that of the $z_{\rm phot,BB}$ case (0.043) by a factor of 2.4,
in agreement with the previous suggestion that the $\zphot$ determination could be improved with the inclusion of medium-band data.
Our $\zphot$ estimation method with the medium-band data opens up a possibility of constructing QLFs at redshift bins finer than previous attempts using broad-band-based $\zphot$ where they constructed QLFs with a coarse bin (e.g., $4.7<z<5.4$ in \citetalias{McGreer18}).

In summary, using the medium-band data, we can estimate the $\zphot$ values of quasars accurately,
comparable to the low-resolution spectroscopy.
As we described above, the $\zphot$ values of high redshift quasars with $i<23$ mag determined by the broad- and medium-band data are reasonably matched to $\zspec$ by an uncertainty of $\langle |\Delta z |/ (1+z) \rangle = 0.016$.
Together with the low contamination rate of our medium-band-based approach, a percentage-level $\zphot$ accuracy improves the luminosity function and the number density estimation of $z\sim5$ quasars and can even allow us to trace large scale distribution of quasars.

The amount of on-source integration we spent on each object ($i<23$ mag) was about 2 to 3 hours. This was for using a 2.1 m telescope under the seeing of $1\farcs0$ to $1\farcs5$.
In comparison, for the spectroscopic observations with Gemini or Magellan, we invested about 1-2 hours of time per target, including overheads.
Considering that 1-2 m class telescope time is much more readily available, the medium-band-based approach is a very cost-effective way to identify high redshift quasars and measure their redshifts to 1-2 \% accuracy.

\section{Implication on the QLF at $\lowercase{z}\sim5$} \label{sec:implication}

Among the newly discovered 10 quasars, three quasars, IMS J021315$-$043341, IMS J021811$-$064843, and IMS J220635$+$020136, were not reported in the final sample of \citetalias{McGreer18}
even as quasar candidates, though these quasars are located in their survey area.
The main difference in the broad-band selection between ours and \citetalias{McGreer18} is the presence of the NIR data from IMS, so this could be a reason for us picking up new quasars in the area already surveyed by \citetalias{McGreer18}.
As shown in the middle panel of Figure \ref{fig:ccd} and Table \ref{tbl:broad}, however, their $riz$ colors (the colors used by \citetalias{McGreer18} for quasar selection) are quite ordinary to be selected as quasar candidates.
Also, they are not particularly faint ($i < 22.4$ mag) to be missed due to large photometry uncertainties.
Another possible reason for the rejection is the stellar source classification of \citetalias{McGreer18} by using the difference of PSF-matched magnitude ($i_{\rm PSF}$) and AUTO magnitudes ($i_{\rm AUTO}$) in $i$-band; $i_{\rm AUTO}-i_{\rm PSF}>-0.15$ mag, but the quasars also satisfy this criterion.
Overall, the three quasars deserve to be selected by \citetalias{McGreer18} even without the NIR data, but they are not.
The differences in photometry between \citetalias{McGreer18} and this work may be the reason, like the four \citetalias{McGreer18} quasars excluded from our candidates (see Section \ref{sec:coleff}).
But we could not verify this because of the lack of the full catalog of \citetalias{McGreer18} in our hand.

We estimated the chance of finding these quasars from the selection functions from \citetalias{McGreer18}.
Based on the spectral properties ($z$, $M_{1450}$, $\alpc$, and EW in Table \ref{tbl:fit}),  the probabilities of finding the three quasars are as high as $\sim95~\%$,
meaning that the quasars are not outliers.
We can update the binned QLF of \citetalias{McGreer18} by the three quasars in their sample.
Assuming the same photometric (94 \%) and spectroscopic (86 \%) completeness of \citetalias{McGreer18} for the three quasars ($21.46 < i < 22.35$ mag),
the number counts corrected by the incompleteness ($N_{cor}$ in Table 1 in \citetalias{McGreer18}) in the magnitude bins of $M_{1450}=-24.35$ and $-23.65$ mag increase from 18.0 and 7.8 to 20.6 and 9.1, respectively,
corresponding to the increase in the binned QLF values at the faint-end by 15 \%.
This is a modest increase and is consistent with the results from \citetalias{McGreer18} within the error.
Yet, the discovery of the three new quasars in the previously surveyed area suggests the importance of independent surveys and applying different methods to gain a complete sample of high redshift quasars.

Our results of finding $z\sim5$ quasars support the scenario of the minor contribution of quasars to the cosmic re-ionization, as the studies of high redshift quasars have suggested so far (e.g., \citealt{Willott10,Kim15,Kashikawa15,Onoue17}; \citetalias{McGreer18}).
Several tens of candidates remain to be observed with the medium-bands, and the ionizing emissivity by quasars at the faint magnitude range of $M_{1450}\sim-23$ mag could change with our future sample with medium-band observations.
However, even if we adopt a pessimistic identification rate of 53 \% (based on the $22 < i < 23$ mag quasar sample) for these remaining faint quasar candidates, the expected binned QLF at $z\sim5$ is marginally in line with the $3\sigma$ upper limit by \citetalias{McGreer18}, meaning that faint quasars contribute to a minor fraction of UV photons to ionize IGM.
The gap in the $z \sim 5$ quasar number density between optical and X-ray surveys would still remain unsolved.

\section{SUMMARY \label{sec:summary}}

We have performed a $z\sim5$ quasar survey with a medium-band-based approach to improve faint quasar candidate selection based on the broad-band colors.
The follow-up imaging and spectroscopy allow us to find ten new quasars at $z\sim5$, among which three were missed in the surveys covering the same area.
Using medium-band data of 35 spectroscopically identified quasars, we demonstrate that quasars can be distinguished effectively from other objects (e.g., brown dwarfs, and galaxies) by imposing medium-band selection criteria to the broad-band selected candidates ($\gtrsim20$ \% of broad-band selected sample are ruled out).
Furthermore, with the inclusion of the medium-band data, the $\zphot$ accuracy improves by a factor of 2-3 in comparison to $z_{\rm phot,BB}$, producing a nearly 1\% level accuracy of $\langle |\Delta z|/(1+z) \rangle =0.016$ (or $\sigma_{\rm NMAD}=0.016$).
Despite our discovery of new faint quasars, the scarcity of $z\sim5$ quasars is consistent with the recent suggestions that the high redshift quasars are not main contributors to the cosmic re-ionization in the early universe.
Based on the high accuracy of the $\zphot$ determination, we expect that the completion of the medium-band survey will enable us to improve the constraint on the faint-end slope of the QLF at $z\sim5$ in the near future.

\acknowledgments

This work was supported by the National Research Foundation of Korea (NRF) grant, 
No. 2017R1A3A3001362, funded by the Korea government (MSIP).
This work was supported by K-GMT Science Program (PID:GS-2016B-Q-11, GS-2016B-Q-46, GS-2017A-Q-19, GS-2018A-Q-220, and GN-2018A-Q-315) of Korea Astronomy and Space Science Institute (KASI).
Based on observations obtained with MegaPrime/MegaCam, a joint project of CFHT and CEA/IRFU, at the Canada-France-Hawaii Telescope (CFHT) which is operated by the National Research Council (NRC) of Canada, the Institut National des Science de l'Univers of the Centre National de la Recherche Scientifique (CNRS) of France, and the University of Hawaii. 
This work is based in part on data products produced at Terapix available at the Canadian Astronomy Data Centre as part of the Canada-France-Hawaii Telescope Legacy Survey, a collaborative project of NRC and CNRS.
The United Kingdom Infrared Telescope (UKIRT) is supported by NASA and operated under an agreement among the University of Hawaii, the University of Arizona, and Lockheed Martin Advanced Technology Center; operations are enabled through the cooperation of the Joint Astronomy Centre of the Science and Technology Facilities Council of the U.K.
Based on observations obtained at the Gemini Observatory acquired through the Gemini Science Archive and processed using the Gemini IRAF package, which is operated by the Association of Universities for Research in Astronomy, Inc., under a cooperative agreement with the NSF on behalf of the Gemini partnership: the National Science Foundation 
(United States), the National Research Council (Canada), CONICYT (Chile), the Australian Research Council (Australia), Minist\'{e}rio da Ci\^{e}ncia, Tecnologia e Inova\c{c}\~{a}o (Brazil) and Ministerio de Ciencia, Tecnolog\'{i}a e Innovaci\'{o}n Productiva (Argentina).
This paper includes data gathered with the 6.5 meter Magellan Telescopes located at Las Campanas Observatory, Chile.
This research was supported by Basic Science Research Program through the National Research Foundation of Korea (NRF) funded by the Ministry of Education (NRF-2017R1A6A3A04005158).
This paper includes data taken at The McDonald Observatory of The University of Texas at Austin.

\vspace{5mm}
\facilities{UKIRT (WFCAM), Struve (SQUEAN), Magellan:Baade (IMACS), Gemini:South (GMOS-S), Gemini:Gillett (GMOS-N),  CFHT (MegaCam), Sloan }

\software{SExtractor \citep{Bertin96}, SWarp \citep{Bertin10}}




\appendix

\section{CORRECTION FOR THE BROAD-BAND COLORS} \label{sec:coffset}

In this section, we describe how we calculate the color offsets of each CFHTLS tile to improve the color selection in this work.
We used the median stellar loci of 0.3 million SDSS-2MASS stars of \cite{Covey07} as a reference.
Though their colors are not corrected for the Galactic extinction, the shape of the loci is consistent with the recent loci based on the 1 million SDSS-2MASS-WISE stars with a low extinction of $A_{r}<0.125$ \citep{Davenport14}.
Furthermore, the loci of \cite{Covey07} are also in line with those of \cite{Gwyn12} based on the point sources in CFHTLS data.
Note that we used the loci of \cite{Covey07}, instead of \cite{Davenport14} that used larger color bins in extreme cases (e.g., $r-i>2$).
For the objects classified as star (CLASS\_STAR $>0.95$ from \texttt{SExtractor}) within the magnitude range of $17<r<21$
in each CFHTLS tile, 
we estimated the color offsets $C^{\rm offset}_{k}$ (where the index $k$ indicates the color; $g-r$, $r-i$, $i-z$ and $i-J$),
which minimize the color distance factor $D_{\rm color}$, given as

\begin{equation}
(D_{\rm color})^{2}=\sum_{i} \sum_{k} \frac{(X^{\rm obj}_{i,k}  - X^{\rm locus}_{i,k} + C^{\rm offset}_{k} )^{2}}{( \sigma^{\rm obj}_{i,k} )^{2} + ( \sigma^{\rm locus}_{i,k} )^{2}} \label{eq:distance},
\end{equation}

\noindent where  $X^{\rm obj}_{i,k}$ is the $k$ color value of the $i$-th object, 
$X^{\rm locus}_{i,k}$ is the $k$ color value of the nearest stellar locus of \cite{Covey07} to $X^{\rm obj}_{i,k}$,
$\sigma^{\rm locus}_{i,k}$ is the quadratic sum of magnitude errors consisting the $k$ color of the $i$-th object,
and $\sigma^{\rm locus}_{i,k}$ is the given error of $X^{\rm locus}_{i,k}$ by \cite{Covey07}.
For the whole survey area, the mean values of $C^{\rm offset}_{k}$ are less than 0.2 mag with small standard deviations of $\sim 0.05$ mag; 
$C^{\rm offset}_{g-r}=-0.02\pm0.05$, $C^{\rm offset}_{r-i}=-0.01\pm0.05$, $C^{\rm offset}_{i-z}=-0.07\pm0.06$,
and $C^{\rm offset}_{i-J}= 0.18\pm0.04$.
The $C^{\rm offset}_{i-J}$ are much larger than the other $C^{\rm offset}_{k}$ in average,
indicating that the $J$-band magnitudes might be slightly over estimated when we introduce the bright 2MASS stars for the $zp$ estimation of IMS data.
We list the $C^{\rm offset}_{k}$ values of our candidates with spectroscopy data in Table \ref{tbl:coffset}.

\section{SPECTRA OF NON-QUASAR OBJECTS\label{sec:nonqso}}

\begin{figure*}
\centering
\epsscale{1.0}
\plotone{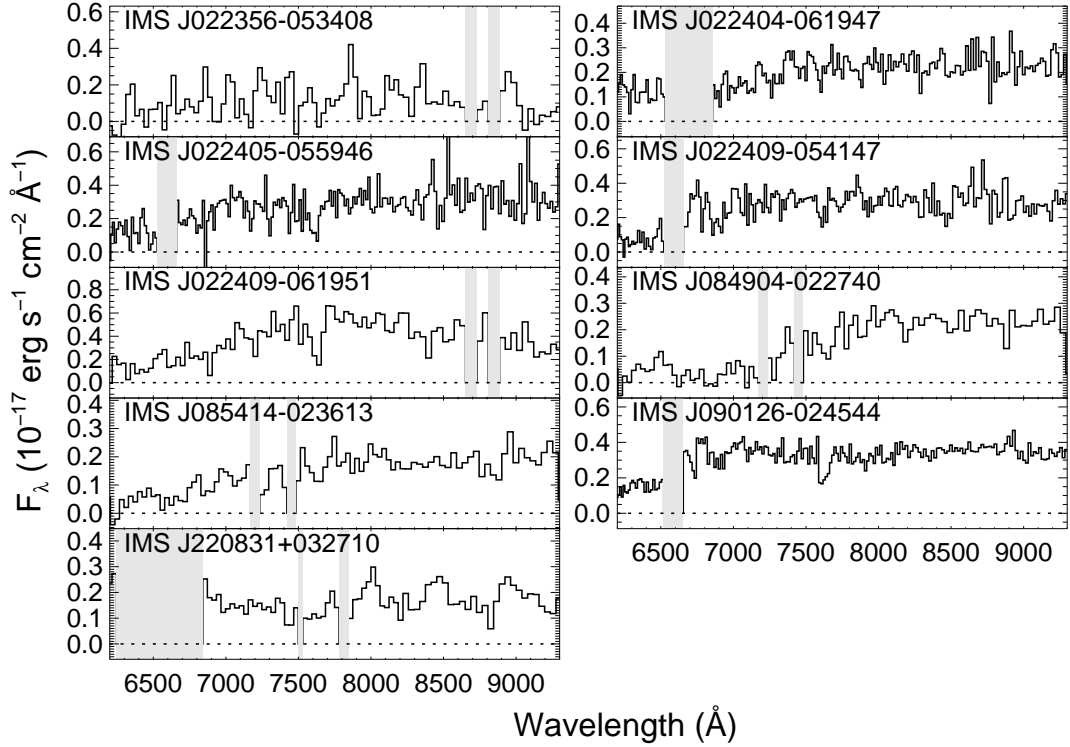}
\caption{
The optical spectra of non-quasar objects. The binned spectra are shown as the black solid lines.
The dotted lines indicate $F_{\lambda}=0$ and the shaded regions represent the bad column area on CCD.
\label{fig:nonqso}}
\end{figure*}

As we described in Section \ref{sec:image}, spectroscopic data were obtained for some of the broad-band-selected quasar candidates before we improved our photometry.
Later, these were excluded from quasar candidates based on the improved broad-band photometry.
Not surprisingly, these objects were spectroscopically identified as non-quasars.
This section provides spectra of these non-quasar objects.
The spectroscopic observations of these objects were carried out with GMOS on the Gemini North/South 8 m Telescopes (PID:GS-2016B-Q-46, GS-2017A-Q-19, and GN-2018A-Q-315) and IMACS on the Magellan Baade 6.5 m Telescope.
The information of the observing runs and their $i$-band magnitudes are listed in Table \ref{tbl:nonqso}, 
and Figure \ref{fig:nonqso} shows their optical spectra.
These candidates are identified as non-quasar objects without any break or line feature at $\gtrsim 7000 \rm \AA$ as we saw for our newly discovered quasars.
The spectra obtained with IMACS show increased fluxes at $\sim6600~\rm\AA$ since it is close to the CCD gap.
However, there is a significant continuum emission at $\lambda_{\rm obs} <6500~\rm\AA$ with no emission line features in both the 1D and the 2D spectra.
Therefore, these objects are regarded as non-quasar objects.

\begin{deluxetable*}{lrrrr}
\tabletypesize{\scriptsize}
\tablecaption{Color Offsets of Spectroscopically Identified Candidates for $z\sim5$ Quasars\label{tbl:coffset}}
\tablewidth{0pt}
\tablehead{
\colhead{ID} & \colhead{$C^{\rm offset}_{g-r}$} & \colhead{$C^{\rm offset}_{r-i}$} & \colhead{$C^{\rm offset}_{i-z}$} & \colhead{$C^{\rm offset}_{i-J}$} \\
 & \colhead{(mag)} & \colhead{(mag)} & \colhead{(mag)} & \colhead{(mag)}
}
\startdata
\multicolumn{5}{c}{Spectroscopically identified quasars}\\
IMS J021315$-$043341 & $-0.05$ & $-0.01$ & $-0.08$ & $ 0.20$ \\ 
IMS J021523$-$052946 & $-0.02$ & $ 0.02$ & $-0.06$ & $ 0.21$ \\ 
IMS J021811$-$064843 & $-0.04$ & $-0.01$ & $-0.15$ & $ 0.12$ \\ 
IMS J022112$-$034232 & $-0.04$ & $-0.01$ & $-0.09$ & $ 0.21$ \\ 
IMS J022113$-$034252 & $-0.04$ & $-0.01$ & $-0.09$ & $ 0.21$ \\ 
IMS J085024$-$041850 & $-0.03$ & $ 0.03$ & $-0.05$ & $ 0.13$ \\ 
IMS J085028$-$050607 & $-0.02$ & $ 0.04$ & $ 0.02$ & $ 0.11$ \\ 
IMS J085225$-$051413 & $ 0.00$ & $ 0.03$ & $ 0.02$ & $ 0.11$ \\ 
IMS J085324$-$045626 & $ 0.00$ & $ 0.03$ & $ 0.02$ & $ 0.11$ \\ 
IMS J135747$+$530543 & $ 0.03$ & $-0.04$ & $-0.02$ & $ 0.19$ \\ 
IMS J135856$+$514317 & $-0.04$ & $ 0.06$ & $-0.06$ & $ 0.23$ \\ 
IMS J140147$+$564145 & $-0.01$ & $ 0.06$ & $-0.10$ & $ 0.16$ \\ 
IMS J140150$+$514310 & $-0.04$ & $ 0.06$ & $-0.06$ & $ 0.23$ \\ 
IMS J140440$+$565651 & $-0.02$ & $ 0.07$ & $-0.03$ & $ 0.17$ \\ 
IMS J141432$+$573234 & $ 0.04$ & $ 0.02$ & $-0.04$ & $ 0.21$ \\ 
IMS J142635$+$543623 & $ 0.01$ & $ 0.01$ & $-0.10$ & $ 0.12$ \\ 
IMS J142854$+$564602 & $-0.01$ & $ 0.00$ & $-0.08$ & $ 0.23$ \\ 
IMS J143156$+$560201 & $-0.01$ & $ 0.00$ & $-0.08$ & $ 0.23$ \\ 
IMS J143705$+$522801 & $ 0.00$ & $ 0.01$ & $-0.08$ & $ 0.23$ \\ 
IMS J143757$+$515115 & $-0.03$ & $-0.03$ & $-0.04$ & $ 0.20$ \\ 
IMS J143804$+$573646 & $ 0.00$ & $ 0.01$ & $-0.01$ & $ 0.26$ \\ 
IMS J143831$+$563946 & $ 0.02$ & $-0.07$ & $-0.02$ & $ 0.27$ \\ 
IMS J143945$+$562627 & $ 0.02$ & $-0.07$ & $-0.02$ & $ 0.27$ \\ 
IMS J220233$+$013120 & $-0.05$ & $-0.03$ & $-0.09$ & $ 0.18$ \\ 
IMS J220522$+$025730 & $-0.05$ & $-0.04$ & $-0.12$ & $ 0.09$ \\ 
IMS J220635$+$020136 & $ 0.02$ & $-0.01$ & $-0.13$ & $ 0.12$ \\ 
IMS J221004$+$025424 & $-0.02$ & $ 0.03$ & $ 0.03$ & $ 0.13$ \\ 
IMS J221037$+$024314 & $-0.02$ & $ 0.03$ & $ 0.03$ & $ 0.13$ \\ 
IMS J221118$+$031207 & $-0.02$ & $ 0.03$ & $ 0.03$ & $ 0.13$ \\ 
IMS J221251$-$004231 & $-0.10$ & $-0.01$ & $-0.09$ & $ 0.16$ \\ 
IMS J221310$-$002428 & $-0.10$ & $-0.01$ & $-0.09$ & $ 0.16$ \\ 
IMS J221520$-$000908 & $-0.04$ & $-0.15$ & $ 0.04$ & $ 0.17$ \\ 
IMS J221622$+$013815 & $-0.04$ & $-0.01$ & $-0.09$ & $ 0.14$ \\ 
IMS J221644$+$001348 & $-0.04$ & $-0.07$ & $-0.10$ & $ 0.12$ \\ 
IMS J222216$-$000406 & $-0.04$ & $-0.05$ & $-0.11$ & $ 0.12$ \\ 
\hline
\multicolumn{5}{c}{Spectroscopically identified non-quasars}\\
IMS J022525$-$044642 & $ 0.00$ & $ 0.05$ & $-0.10$ & $ 0.22$ \\ 
IMS J090540$-$011038 & $ 0.00$ & $-0.02$ & $-0.10$ & $ 0.15$ \\ 
\enddata
\end{deluxetable*}

\begin{deluxetable*}{ccccc}
\tabletypesize{\scriptsize}
\tablecaption{Spectroscopic Observations of Non-quasar objects\label{tbl:nonqso}}
\tablewidth{0pt}
\tablehead{
\colhead{ID} & \colhead{Telescope/instrument} & \colhead{Date} & \colhead{Exposure time (s)} & \colhead{$i$ (mag)}
}
\startdata
IMS J022356$-$053408 & Gemini/GMOS-S  & 2016 Sep 3-4 & 5760 & 22.79 \\
IMS J022404$-$061947 & Magellan/IMACS & 2016 Dec 5   & 3600 & 22.41 \\
IMS J022405$-$055946 & Magellan/IMACS & 2016 Dec 6   & 1800 & 22.09 \\
IMS J022409$-$054147 & Magellan/IMACS & 2016 Dec 6   & 1800 & 22.05 \\
IMS J022409$-$061951 & Gemini/GMOS-S  & 2016 Sep 3   & 960  & 21.56 \\
IMS J084904$-$022740 & Gemini/GMOS-S  & 2017 Feb 22  & 4800 & 22.71 \\
IMS J085414$-$023613 & Gemini/GMOS-S  & 2017 Feb 22  & 4800 & 22.76 \\
IMS J090126$-$024544 & Magellan/IMACS & 2016 Dec 6   & 2100 & 21.91 \\
IMS J220831$+$032710 &  Gemini/GMOS-S & 2018 Jun 22  & 3000 & 22.84 \\
\enddata
\tablecomments{These objects were selected before the improved photometry described in Section \ref{sec:image}.}
\end{deluxetable*}



\end{document}